\documentclass[aps,prb,letterpaper,superscriptaddress,twocolumn,showpacs,floatfix,10pt]{revtex4-1}

\usepackage{amsfonts}
\usepackage{amsmath}
\usepackage{dsfont}
\usepackage{amssymb}
\usepackage{graphicx}
\usepackage{epstopdf}
\usepackage{epsfig}
\usepackage{times}
\usepackage{color}
\newcommand{\be}{\begin{equation}}
\newcommand{\ee}{\end{equation}}
\newcommand{\bea}{\begin{eqnarray}}
\newcommand{\eea}{\end{eqnarray}}
\newcommand{\ba}{\begin{array}}
\newcommand{\ea}{\end{array}}

\newcommand{\ket}[1]{|#1\rangle}
\newcommand{\bra}[1]{\langle #1|}

\newcommand{\cC}{{\cal{C}}}

\newcommand{\cT}{{\cal{T}}}

\begin{document}
\title{`Gauging' time reversal symmetry in tensor network states}
\author{Xie Chen}
\affiliation{Department of Physics and Institute for Quantum Information and Matter, California Institute of Technology, Pasadena, CA 91125, USA}
\affiliation{Department of Physics, University of California, Berkeley, CA, 94720, USA}
\author{Ashvin Vishwanath}
\affiliation{Department of Physics, University of California, Berkeley, CA, 94720, USA}

\begin{abstract}
It is well known that unitary symmetries can be `gauged', i.e. defined to act in a local way, which leads to a corresponding gauge field. Gauging, for example, the charge conservation symmetry leads to electromagnetic gauge fields. It is an open question whether an analogous process is possible for time reversal which is an anti-unitary symmetry. Here we discuss a route to gauging time reversal symmetry which applies to gapped quantum ground states that admit a tensor network representation. The tensor network representation of quantum states provides a notion of locality for the wave function coefficient and hence a notion of locality for the action of complex conjugation in anti-unitary symmetries. Based on that, we show how time reversal can be applied locally and also describe time reversal symmetry twists which act as gauge fluxes through nontrivial loops in the system.  As with unitary symmetries, gauging time reversal provides useful access to the physical properties of the system. We show how topological invariants of certain time reversal symmetric topological phases in $D=1,2$ are readily extracted using these ideas. 
\end{abstract}

\maketitle


\section{Introduction}

For condensed matter systems with global symmetry, coupling to the corresponding gauge field provides a useful access to the physical properties of the system. For example, in systems with charge conservation ($U(1)$) symmetry, coupling to the electromagnetic field and measuring the induced charge or current is a direct probe of the low energy excitations of the system. In gapped systems without low energy excitations, coupling to gauge field and introducing gauge fluxes to the system creates finite energy excitations which can reveal important information about the topological order of the system. For example, in fractional quantum Hall systems with $U(1)$ symmetry, inserting a magnetic flux $\Phi$ results in the accumulation of charge $\sigma_{xy}\Phi$ around the flux and hence is a direct measure of the quantized Hall conductance.\cite{Laughlin1981} When the continuous $U(1)$ symmetry is broken down to a discrete symmetry, discrete fluxes can be introduced. For example, in superconductors where the $U(1)$ symmetry breaks down to $Z_2$, fluxes in multiples of $hc/2e$ can penetrate the system and one of the most important properties of the topological $p+ip$ superconductor is that each $hc/2e$ flux contains a Majorana zero mode.\cite{Read2000} Similarly, systems with nonabelian symmetries can be coupled to nonabelian gauge fields.\cite{Yang1954} In generally, coupling topological phases with various symmetries (symmetry protected topological (SPT) phases or symmetry enriched topological (SET) phases) to the corresponding gauge field results in nontrivial responses (like nontrivial statistics, Hall conductance, symmetry fractionalization, etc.) and provides an important tool in distinguishing these phases.\cite{Levin2012,Lu2012,Gu2014,Cheng2014,Lu2013,Liu2013,Hung2013}

Can time reversal symmetry be similarly gauged? 
The anti-unitary nature of time reversal symmetry has made such a notion hard to define and we can see how a straight-forward generalization from unitary symmetries fails. In coupling systems with global unitary symmetries to gauge fields, we first find the action of the symmetry on the local Hilbert spaces in the system. For $U(1)$ symmetry it would be adding phase factors on the local charges and for $SU(2)$ symmetry it would be local rotation of spins. Then extra degrees of freedom -- the gauge field -- are introduced into the system which transform under the local symmetry and couple to the original degrees of freedom in such a way that the total system is now invariant under arbitrary local actions of the symmetry. However, when trying to implement the same procedure for time reversal symmetry, we fail at the first step -- we do not know how to define a local action of time reversal! Global time reversal symmetry involves not only unitary transformations on local Hilbert spaces (like inverting spins) but also a complex conjugation operation on the coefficient of each basis state in the wave function .
\be
\ba{ll}
& \cT \sum_{i_1,...,i_N} C_{i_1,...,i_N} \ket{i_1,...,i_N} \\
= & \sum_{i_1,...,i_N} C^*_{i_1,...,i_N} U_1 \otimes ...\otimes U_N \ket{i_1,...,i_N}
\ea
\ee
Therefore, to couple a system to time reversal gauge field, first we need to define how complex conjugation acts on $C_{i_1,...,i_N}$ locally. (The local action of the unitary part is straight-forwardly defined.)

To define such an action, we need to decide if we should consider $C_{i_1,...,i_N}$ to be localized on a particular lattice site or distributed over the entire system? Different ways to divide the coefficient would result in different local actions of time reversal. If we designate $C_{i_1,...,i_N}$ to be localized on site 1, then local action of time reversal on site 1 takes complex conjugation of $C_{i_1,...,i_N}$ while local actions on other sites do not. If we think of $C_{i_1,...,i_N}$ as composed of various parts on different lattice sites, then time reversal on a subsystem takes complex conjugation of the corresponding part and not the others. Then the question is, what is the most meaningful way to do this?

Useful insight can be obtained by thinking of a simple case -- a product state
\be
\ket{\psi}= \prod_k (a^{0_k}\ket{0} + a^{1_k}\ket{1}) = \sum_{i_1,...,i_N} a^{i_1}...a^{i_N}\ket{i_1,...,i_N}
\ee
In such a product state, it is natural to divide $C_{i_1,...,i_N}$ into $N$ parts $a^{i_k}$, $k=1,...,N$ and associate them with each site $k$. Acting time reversal locally on site $k$ then involves taking complex conjugation on $a^{i_k}$. With this definition of local time reversal symmetry action, we can see that if the total state $\ket{\psi}$ is invariant under global time reversal action, then it is also invariant under local time reversal action (up to a phase factor). This is similar to the action of unitary symmetries on a product symmetric state.

How to divide the wave function coefficient $C_{i_1,...,i_N}$ in a many-body entangled state? The tensor network representation of many-body entangled states\cite{Cirac2009,Vidal2009} provides a very natural way to do so. The tensor network representation describes a many-body wave function in terms of a set of local tensors $T^{i_k}$.
\be
\ket{\psi}=\sum_{i_1,...,i_N} tTr\left(T^{i_1} ... T^{i_N}\right) \ket{i_1,...,i_N}
\ee
where $tTr$ denotes tensor contraction. We can then think of the tensors $T^{i_k}$ as local pieces of the total coefficient $C_{i_1,...,i_N}$ and define local action of complex conjugation on site $k$ as taking complex conjugation of $T^{i_k}$. Combined with the local action of the unitary part of time reversal, we obtain a local way of implementing time reversal symmetry. In this paper, we focus on gapped short range correlated quantum states which can be well described using the tensor network formalism and such a definition of local time reversal action applies.\footnote{In Ref.\onlinecite{Levin2012a} a different notion of local time reversal symmetry action is defined which applies not to ground states but to fractional excitations in topological phases.}

\begin{figure}[htbp]
\begin{center}
\includegraphics[width=8.0cm]{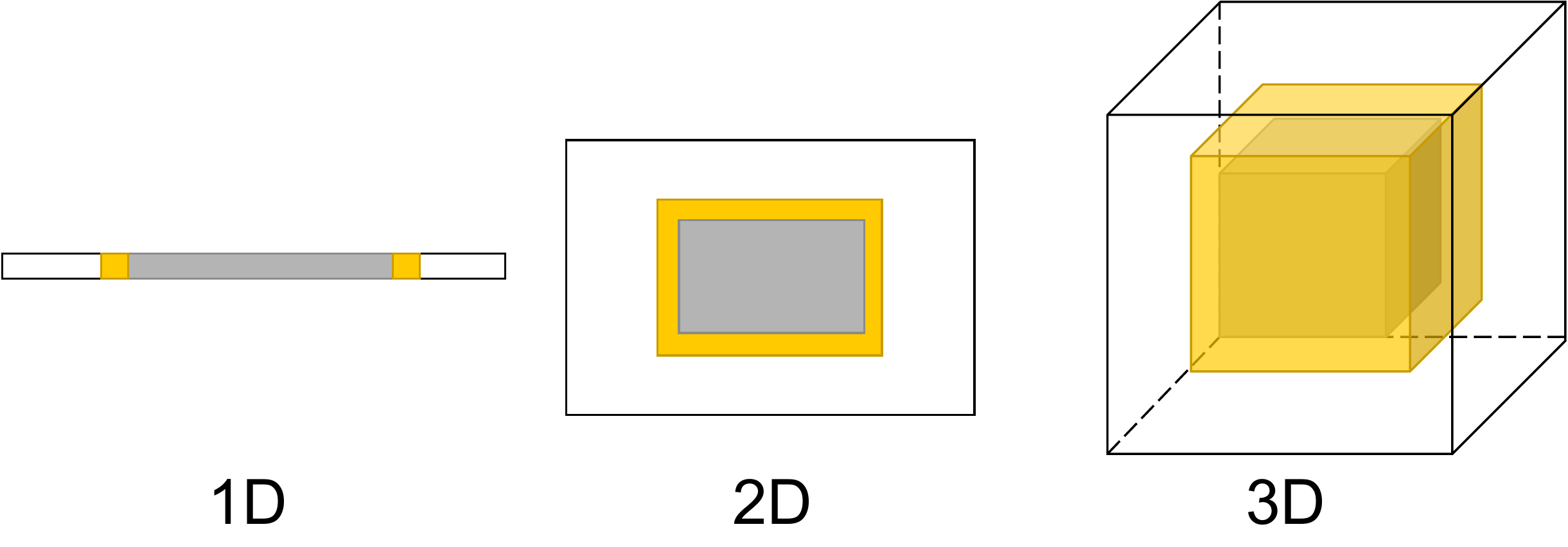}
\caption{Local action of time reversal symmetry (grey+yellow region) on a short range correlated symmetric state only changes the state at the border of the region (yellow region).}
\label{fig:T123}
\end{center}
\end{figure}

Is this a valid and useful definition? First we notice that, such a local action of time reversal leads to similar changes on a gapped symmetric quantum state as that induced by the local action of unitary symmetries. Imagine applying a unitary symmetry to a subregion in a gapped symmetric quantum state. Both deep inside and outside the region, the state should remain invariant. The only change in the state happens at the border of the region of symmetry action, as shown in Fig.\ref{fig:T123}. To see how this can be true in our definition of local time reversal symmetry action on tensor network states, we note that global time reversal symmetry involves complex conjugation on all the tensors $T^{i_k}$ and unitary operations $\prod_k U_k$ on all the physical degrees of freedom. The tensors at site $k$ ($T^{i_k}$) may change into $\tilde{T}^{i_k}$ under complex conjugation and $U_k$. However, if the state is invariant under global time reversal symmetry action, then the changes in $T^{i_k}$ should cancel with that coming from neighboring sites. Because of this, if we apply time reversal symmetry locally (complex conjugation and $U_k$ on sites in a subregion), then tensors both inside and outside the subregion remain effectively invariant while tensors along the border can change. Therefore, intuitively, this definition of local time reversal symmetry action changes the quantum state in a way we would expect.

More concretely, we are going to show in the following sections that using such a definition of local time reversal symmetry action, we can extract topological invariants from a gapped symmetric quantum states and hence identify the symmetry related topological order.

In section \ref{1D}, we discuss the 1D case in terms of matrix product states and present a way to insert time reversal fluxes through a 1D ring. In 1D there are two different time reversal symmetry protected topological (SPT) phases.\cite{Gu2009,Pollmann2010,Pollmann2012} We demonstrate how these two phases can be distinguished from each other using the projective composition rule of time reversal twists induced by the inserted time reversal fluxes. That is, two time reversal twists compose into identity up to a universal phase factor characterizing the underlying SPT order of the state. In 1D, the distinction between different time reversal SPT phases has been well understood in the matrix product formalism.\cite{Pollmann2010,Pollmann2012,Chen2011,Schuch2011} Our discussion is just a reinterpretation of that procedure in terms of local time reversal symmetry action and time reversal twists. 

In section \ref{2D}, we move onto the 2D case, where we define local action of time reversal in tensor product states and discuss how to insert time reversal fluxes through nontrivial loops in the systems. To facilitate discussion, we first review the procedure for the unitary $Z_2$ symmetry and demonstrate how topological invariants of the 2D $Z_2$ SPT phases can be extracted from the projective composition rules of $Z_2$ symmetry twist lines. Then we study a 2D state with trivial time reversal SPT order, a 2D SPT phase with $Z_2\times Z_2^T$ symmetry and a 2D $Z_2$ gauge theory with time reversal symmetry and see how the topological invariants of these phases can be extracted similarly. In particular, we find that for the trivial time reversal SPT state the projective composition rule for the time reversal fluxes are all trivial, as expected. For the $Z_2\times Z_2^T$ SPT state and the $Z_2$ gauge theory with time reversal symmetry, a nontrivial $(-1)$ phase factor can appear which is related to the $\cT^2=-1$ transformation law of the $Z_2$ fluxes in the bulk. In section \ref{sum}, we summarize what we have learned and discuss open problems in gauging time reversal symmetry.

\section{In 1D Matrix Product States}
\label{1D}

The matrix product state (MPS) representation of 1D gapped quantum states provides a natural way to divide the wave function coefficient into local pieces. The matrix product state representation reads
\be
\ket{\psi}= \sum_{i_1,i_2,...,i_N} Tr\left(A^{i_1}A^{i_2}...A^{i_N}\right) \ket{i_1,i_2,...,i_N}
\ee
with $D\times D$ matrices $A^i$. We call the $i$'s the physical indices and the left and right indices of $A$'s the inner indices. Similar terminology is used for tensor product states discussed later. The local action of time reversal symmetry on matrix product states has been discussed extensively in the study of 1D symmetry protected topological phases.\cite{Pollmann2010,Pollmann2012,Chen2011,Schuch2011} Here we review the procedure and discuss the notion of time reversal flux and time reversal twists based on such a formalism.

Suppose that the global time reversal symmetry action is $U\otimes ... \otimes U K$, where $K$ denotes complex conjugation in the $\ket{i}$ basis. The global action of time reversal on a MPS reads
\be
\cT \ket{\psi}=\sum_{i_1,i_2,...,i_N} Tr\left(A^{i_1}A^{i_2}...A^{i_N}\right)^*  U\otimes U ... \otimes U\ket{i_1i_2...i_N}
\ee
Then acting time reversal locally on a single site in the matrix product state changes the matrices to
\be
\tilde{A}^{i} = \sum_{j} U^T_{ij} (A^{j})^* \label{Tmps}
\ee
Note that if this is applied to all sites, then it is equivalent to applying time reversal globally. If the state $\ket{\psi}$ is short range correlated and time reversal symmetric, then the MPS representation satisfies
\be
\tilde{A}^{i} = \sum_{j} U^T_{ij} (A^{j})^* = M A^i M^{-1}
\ee
with an invertible matrix $M$. Therefore, applying time reversal on one site is equivalent to inserting $M$ and $M^{-1}$ on the two sides of $A^i$.

If we apply time reversal to a segment of sites $m$ to $m+n$, then the matrices on all sites remain invariant except for those on site $m$ and $m+n$. On site $m$, the matrices change to
\be
\tilde{A}^i = MA_i
\ee
On site $m+n$, the matrices change into
\be
\tilde{A}^i = A^i M^{-1}
\ee 

For example, consider the dimer state with two spin $1/2$'s per site and the spin $1/2$s' on neighboring sites pair into the singlet state $\ket{01}-\ket{10}$. The MPS representation of the state contains matrices
\be
A^{00}=-\ket{1}\bra{0}, A^{01}=-\ket{1}\bra{1}, A^{10}=\ket{0}\bra{0}, A^{11}=\ket{0}\bra{1} \label{mps_dm}
\ee
If time reversal symmetry acts as
\be
\cT = i\sigma_y \otimes ... \otimes i\sigma_y K
\ee
Then acting time reversal on each site changes the matrices as
\be
\tilde{A}^i = (iY) A^{i} (-iY) \label{TAKLT}
\ee
Here we use $X$, $Y$ and $Z$ to denote Pauli matrices on the inner indices.

It is known that the time reversal SPT order can be extracted from $M$ by \cite{Pollmann2010,Pollmann2012,Chen2011,Schuch2011}
\be
M^*M=\beta = \pm 1
\ee
where $\beta=1$ in the trivial SPT phase and $\beta=-1$ in the nontrivial one (including the dimer state where $M=iY$). Now we are going to reinterpret this in terms of time reversal fluxes and time reversal twists.

With the definition of local time reversal symmetry action, we can discuss how to insert time reversal fluxes through a 1D ring, in analogy to unitary symmetries. Let's first recall how the procedure works for unitary symmetries.

Consider, for example, a system with $U(1)$ symmetry $e^{i\theta n_1} \otimes e^{i\theta n_2} \otimes ... \otimes e^{i\theta n_N}$, where $n_k$ counts the number of $U(1)$ charge on each site. WLOG, consider a Hamiltonian with two-body interactions $H=\sum_k h_{k,k+1}$. Inserting a $\phi$ flux through the one dimensional ring corresponds to changing the Hamiltonian term on the boundary as
\be
h_{N,1} \rightarrow e^{i\phi n_N} h_{N,1} e^{-i\phi n_N}
\ee
For all the SPT and SET phases we are considering in this paper, the ground states satisfy that all the local reduced density matrices are determined by local Hamiltonian terms and not affected by terms far away (this is the TQO-2 condition used in the definition of topological order in Ref.\onlinecite{Bravyi2010}). Therefore the two ground states $\ket{\psi}$ and $\ket{\psi}_{\phi}$ without and with flux should have the following relation:
\begin{enumerate}
\item{Away from the boundary, $\ket{\psi}$ and $\ket{\psi}_{\phi}$ should look the same.}
\item{Near the boundary, $\ket{\psi}_{\phi}$ should look the same as $\ket{\psi}$ with symmetry applied on one side of the boundary, i.e. $e^{i\phi n_m} \otimes ... \otimes e^{i\phi n_N} \ket{\psi}$ or $e^{-i\phi n_1} \otimes ... \otimes e^{-i\phi n_m} \ket{\psi}$ with $1<<m<<N$.}
\end{enumerate}
Here by `look the same' we mean that the two states have the same reduced density matrix locally.
 
For time reversal symmetry, we do not know how to couple the system to fluxes on the Hamiltonian level. However, we can couple the symmetric gapped ground state to time reversal fluxes in a way similar to unitary symmetries. Denote the state not coupled / coupled to a time reversal flux as $\ket{\psi}$ and $\ket{\psi}_{\cT}$. We expect that
\begin{enumerate}
\item{Away from the boundary, $\ket{\psi}$ and $\ket{\psi}_{\cT}$ should look the same.}
\item{Near the boundary, $\ket{\psi}_{\cT}$ should look the same as $\ket{\psi}$ with time reversal applied on one side of the boundary, i.e. from site $m$ to $N$ or from site $1$ to $m$ with $1<<m<<N$.}
\end{enumerate}

\begin{figure}[htbp]
\begin{center}
\includegraphics[width=6.0cm]{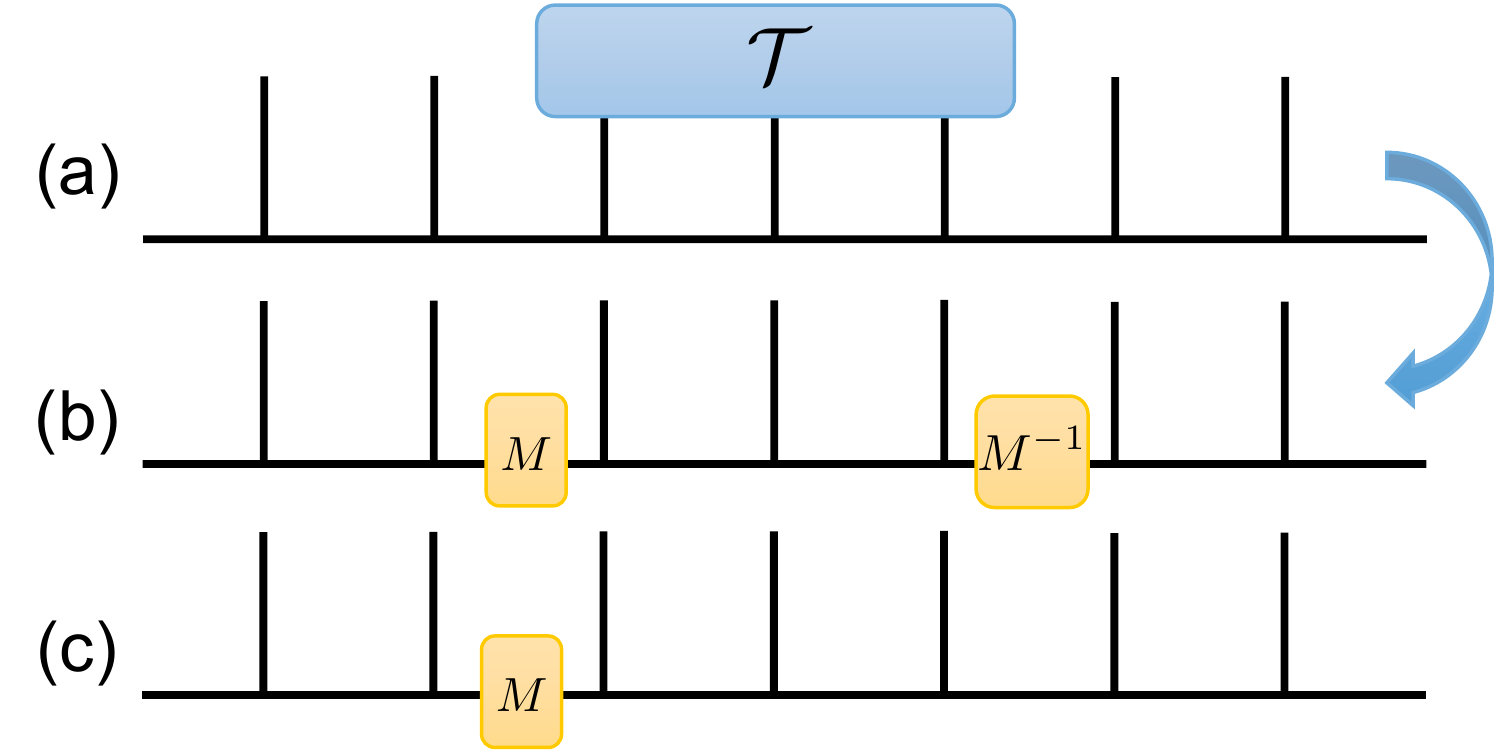}
\caption{(a) Local time reversal symmetry action on a symmetric MPS (b) changes only matrices near the two ends; (c) Coupling an MPS to a time reversal flux corresponds to changing the matrices only at one place. Vertical links represent physical indices of the MPS and horizontal links represent the inner indices of the MPS.}
\label{MPS_flux}
\end{center}
\end{figure}

As we discussed above, in short range correlated matrix product states, local time reversal symmetry action changes the representing matrices only near the two ends of the local region, as shown in Fig.\ref{MPS_flux} (a) and (b). Therefore, if we change the matrices only at one point, as shown in Fig.\ref{MPS_flux} (c), for example at the boundary of the 1D ring, this would correspond to inserting a time reversal flux through the 1D ring. 

In particular, to insert a time reversal flux, we can change the matrices at site $N$ to
\be
A^i \rightarrow A^i M^{-1}
\ee
Or we can change the matrices at site $1$ to
\be
A^i \rightarrow M A^i
\ee
The resulting state $\ket{\psi}_{\cT}$ indeed has the property discussed above when compared to the original state $\ket{\psi}$ without flux and contains a time reversal symmetry twist on the boundary.

The usefulness of this definition becomes evident when we compose two time reversal symmetry twists and extract universal properties of the SPT order from the procedure. Suppose that we insert two time reversal fluxes by changing the matrices on site $1$ twice. This should be equivalent to a state without time reversal flux. However, as we will see, two time reversal twists may differ from zero twist by an important phase factor. On inserting the first flux, the matrices on site $1$ are changed to $A^i \rightarrow M A^i$. On inserting the second flux, the $A_i$ part undergoes the change $A^i \rightarrow M A^i$ again. Moreover, because we are considering time reversal fluxes, we need to take complex conjugation of the first $M$. Therefore, the total change to $A^i$ on site $1$ is
\be
A^i \rightarrow M^*MA^i = \beta A^i
\ee
with $\beta=M^*M=\pm 1$. 
Therefore, the composition of two time reversal twists is equivalent to zero twist up to a phase factor of $\beta$. From this projective composition rule of time reversal twists, we recover the topological invariant $\beta$ characterizing the SPT order of the state. The result remains the same if we insert flux by changing the matrices on site $N$ as $A^i \rightarrow A^i M^{-1}$. 

This corresponds exactly to the procedure of extracting SPT order from the MPS representation of a gapped symmetric state. \cite{Pollmann2010,Pollmann2012,Chen2011,Schuch2011} Here we are merely reinterpreting the procedure as finding the projective composition rule of time reversal twists induced by inserted time reversal fluxes.
It follows from previous discussions that the topological invariant extracted ($\beta$) is independent of the gauge choice we make for the matrix product representation. In particular, we can change the gauge choice of the MPS representation by an invertible matrix $N$. Then inserting a single flux corresponds to changing the matrices on site $1$ by $N^*MN^{-1}$ and
\be
(N^*MN^{-1})^*(N^*MN^{-1}) = M^*M = \beta
\ee
We want to note that this procedure applies to any matrix product state, not only the fixed point ones with zero correlation length as shown in the dimer state example above.

\section{In 2D Tensor Product States}
\label{2D}

Now, we are ready to generalize the procedure to 2D. First we will review in section \ref{Z2} how everything works for unitary symmetries. In particular, we are going to review how the $Z_2$ symmetry acts locally on the tensor product representation of a gapped $Z_2$ symmetric state and how topological invariants of its SPT order can be extracted from the projective composition rules of the $Z_2$ symmetry twist lines. After this preparation, we will move on to 
define how time reversal symmetry acts locally on 2D tensor product states and discuss how to insert time reversal fluxes through nontrivial loops in the system in section \ref{2DT}. In particular, we are going to discuss in detail three examples in sections \ref{2DT}, \ref{2DSPT_Z2Z2T}, \ref{2DSET_Z2T} and demonstrate how topological invariants of the phases can be extracted from the projective composition rules of time reversal twist lines.

\subsection{Review: local action and twists of unitary symmetry in 2D}
\label{Z2}

\begin{figure}[htbp]
\begin{center}
\includegraphics[width=4.0cm]{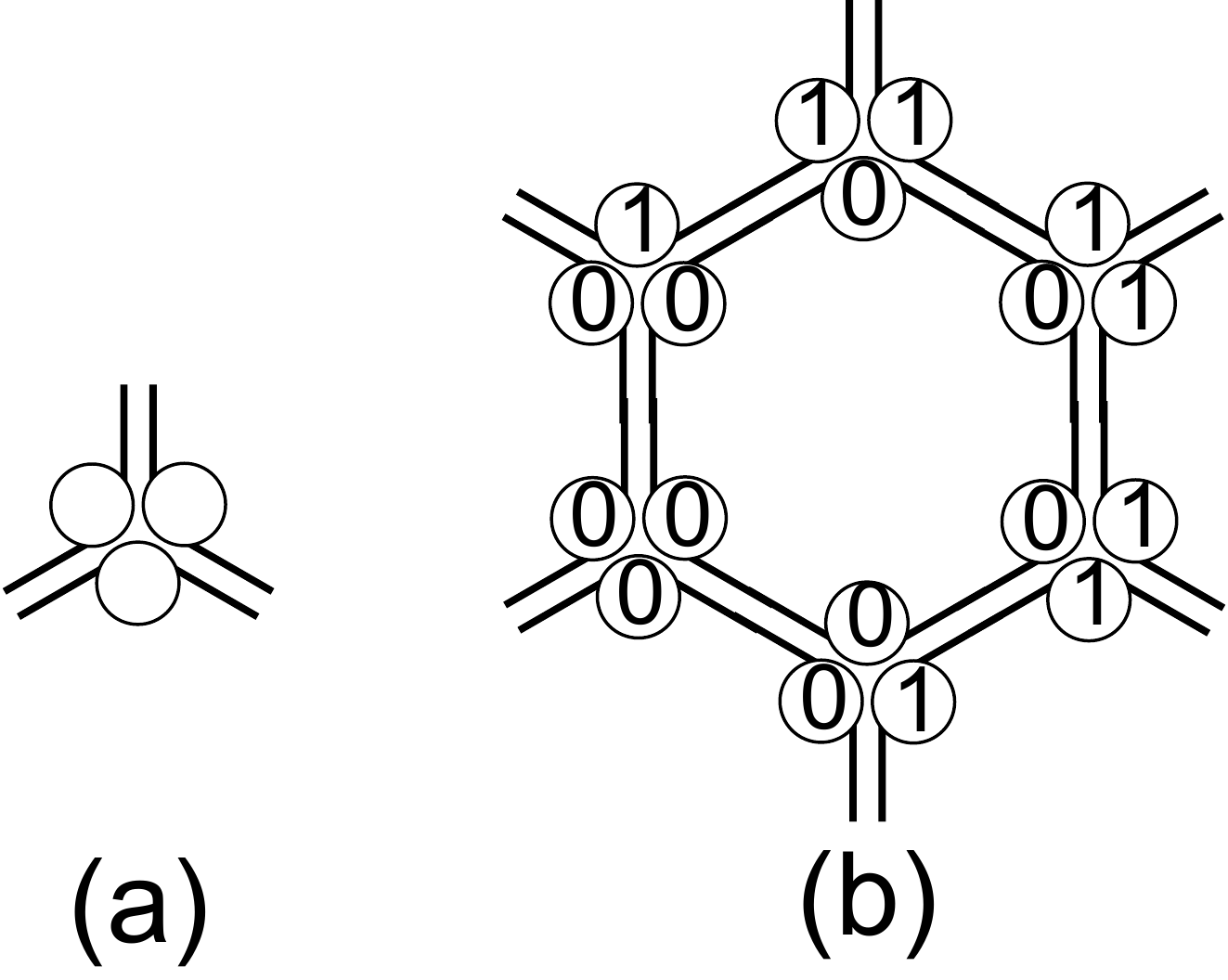}
\caption{Wave function of 2D states on the honeycomb lattice with $Z_2$ SPT order: each lattice site contains three spin $1/2$'s; the six spin $1/2$'s around each plaquette are either all in the $\ket{0}$ state or all in the $\ket{1}$ state. The $Z_2$ symmetry flips between the $\ket{0}$ and $\ket{1}$ state.}
\label{fig:Z2wf}
\end{center}
\end{figure}

Consider the 2D states with trivial and nontrivial $Z_2$ symmetry protected topological order.\cite{Chen2011b,Levin2012,Chen2012a,Chen2013a,Lu2012} We use the form of wave function similar to that in Ref.\onlinecite{Levin2012}. The system lives on a honeycomb lattice, where each lattice site contains three spin $1/2$'s (in state $\ket{0}$ or $\ket{1}$), as shown in Fig.\ref{fig:Z2wf} with three circles. In the ground state, the six spin $1/2$'s around a plaquette are either all in the $\ket{0}$ state or all in the $\ket{1}$ state, forming $Z_2$ domains. A state with trivial $Z_2$ SPT order can be obtained as an equal weight superposition of all $Z_2$ domain configurations 
\be
\ket{\psi_0}  = \sum_{\cC} \ket{\cC} \label{Z2wf0}
\ee
where $\cC$ denotes $Z_2$ domain configurations. A state with nontrivial $Z_2$ SPT order takes the form
\be
\ket{\psi_1}  = \sum_{\cC} (-)^{N_{\cC}}\ket{\cC} \label{Z2wf1}
\ee
where $N_{\cC}$ counts the number of domain wall loops in the configuration. Obviously, these two wave functions are both symmetric under the $Z_2$ symmetry action of flipping $Z_2$ domains $\ket{0} \leftrightarrow \ket{1}$. However, they contain different SPT orders which is reflected in the different statistics of their $Z_2$ fluxes once the symmetry is gauged. The $Z_2$ fluxes have bosonic statistics in the trivial phase represented by $\ket{\psi_0}$ and semionic statistics in the nontrivial phase represented by $\ket{\psi_1}$.\cite{Levin2012} We will see how this difference in SPT order is reflected in the projective composition rule of the $Z_2$ symmetry twist lines as discussed below.

It is particularly helpful to use the tensor network representation of the states and see how the tensors transform under the symmetry. For unitary symmetries, the discussion can be carried out without referring to the tensor network representation. But here we use such a representation because 1. it provides a nice picture of how a state changes under local action of symmetry and with the insertion of symmetry fluxes; 2. it paves the way for our discussion of time reversal symmetry which necessarily depends on a tensor network representation. 

\begin{figure}[htbp]
\begin{center}
\includegraphics[width=7.0cm]{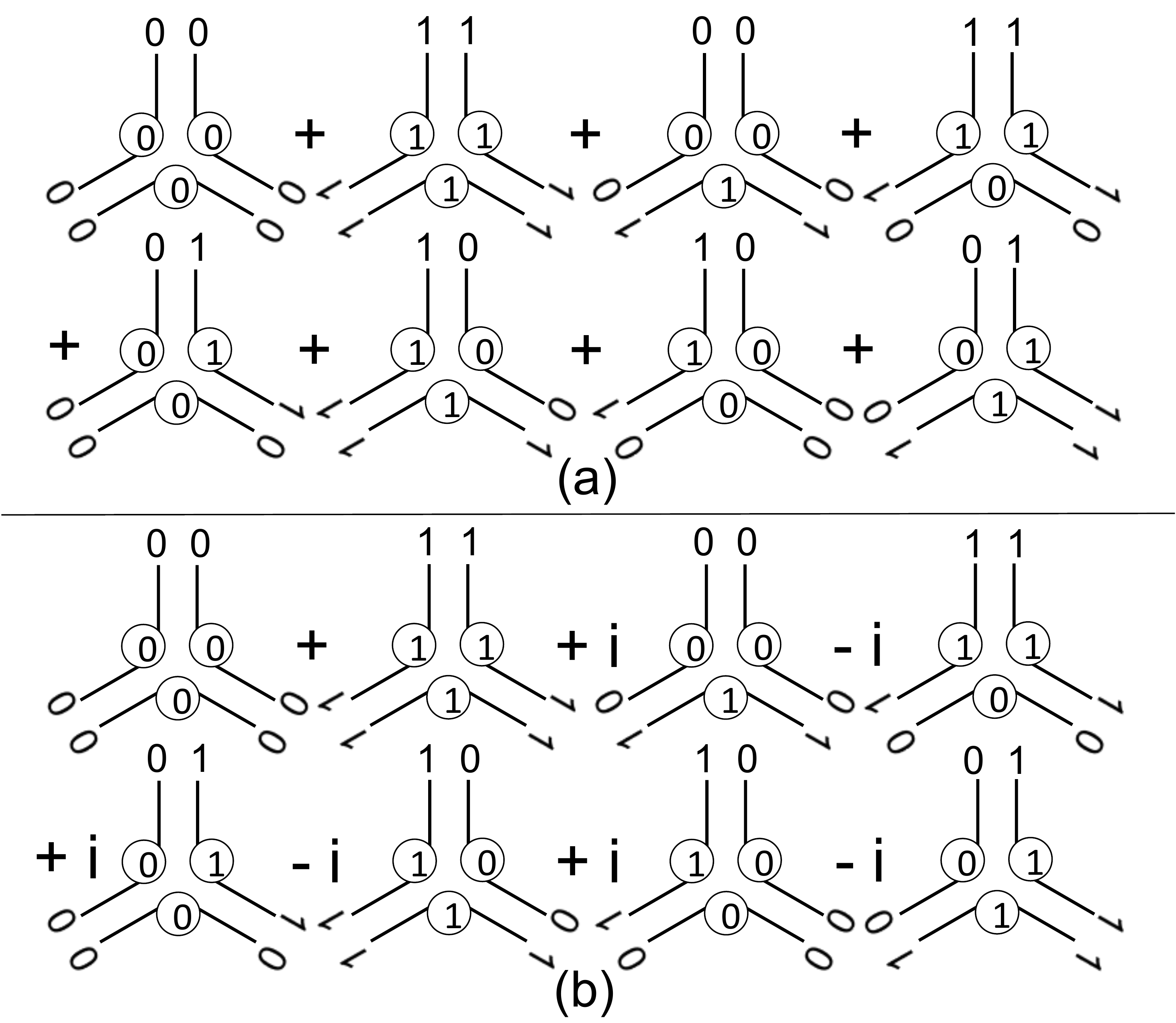}
\caption{Tensors representing the trivial $Z_2$ SPT wave function in Eq. \ref{Z2wf0} and the nontrivial $Z_2$ SPT wave function in Eq.\ref{Z2wf1}. The labels in circles are physical indices and the labels at the end of the links are inner indices.}
\label{fig:Z2_TPS}
\end{center}
\end{figure}

The trivial SPT state $\ket{\psi_0}$ can be represented with tensors given in Fig.\ref{fig:Z2_TPS} (a) while the nontrivial SPT state $\ket{\psi_1}$ can be represented with tensors given in Fig.\ref{fig:Z2_TPS} (b). Note that a physical index and the two inner indices connected to it are always in the same state ($\ket{0}$ or $\ket{1}$). Therefore, when the tensors are contracted together, all physical spins around the same plaquette are in the same state as shown in Fig.\ref{fig:Z2wf}, forming $Z_2$ domains. The tensor for $\ket{\psi_1}$ is similar to the tensor product representation of the double semion state given in Ref.\onlinecite{Gu2009b}. The tensors are the same on A, B sub-lattices.
We can see that the tensor network representation for $\ket{\psi_0}$ can be obtained from that for $\ket{\psi_1}$ simply by removing the phase factors $i$ and $-i$.

\begin{figure}[htbp]
\begin{center}
\includegraphics[width=8.5cm]{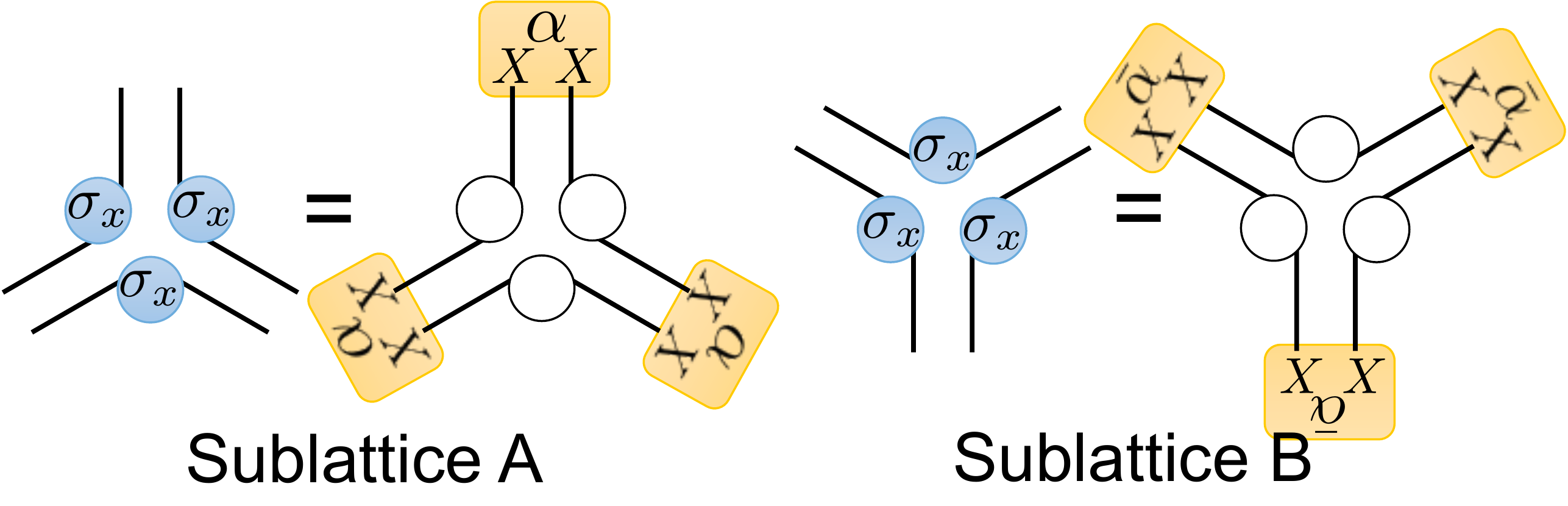}
\caption{Local $Z_2$ symmetry transformation on the tensors representing the state in Eq.\ref{Z2wf0} and \ref{Z2wf1}.}
\label{fig:sym_Z2TPS}
\end{center}
\end{figure}

It is interesting to see how the $Z_2$ symmetry acts locally on the tensors. Obviously the tensors for $\ket{\psi_0}$ and $\ket{\psi_1}$ are not invariant under local action of $Z_2$ symmetry $\ket{0} \leftrightarrow \ket{1}$, but the transformed tensors differ from the original ones by unitary transformations on the inner indices. This relation is shown in Fig.\ref{fig:sym_Z2TPS}, where for $\ket{\psi_0}$ 
\be
\alpha=\bar{\alpha}=1
\ee
and for $\ket{\psi_1}$
\be
\ba{l}
\alpha=\ket{00}\bra{00} + i \ket{01}\bra{01} + i \ket{10}\bra{10} + \ket{11}\bra{11} \\
\bar{\alpha}=\ket{00}\bra{00} - i \ket{01}\bra{01} - i \ket{10}\bra{10} + \ket{11}\bra{11} 
\ea
\label{alpha}
\ee
$\sigma_x$ denotes the spin flip of physical degrees of freedom while $X$ denotes the same operator but on the inner indices. In the following discussion, we call $X \otimes X \alpha$ and  $X \otimes X \bar{\alpha}$ the `inner symmetry operators'. Using the relation given in Fig.\ref{fig:sym_Z2TPS}, it is straight-forward to see that the state is invariant under global $Z_2$ symmetry action and changes only along the border if the $Z_2$ symmetry is applied to a subregion. This is because changes to the inner indices cancel if the index lies within the subregion.

Now we can couple the state to $Z_2$ symmetry fluxes. Consider a $Z_2$ SPT state on a torus. Inserting a flux through a nontrivial loop of the torus results in a symmetry twist line on the torus along the other nontrivial loop. Assume WLOG that the Hamiltonian of the system contains only two-body interactions. Creating symmetry twist lines corresponds to taking all terms in the Hamiltonian $h_{mn}$ that are divided by the twist line and changing them to
\be
h_{mn} \rightarrow \sigma^m_x h_{mn} \sigma^m_x
\ee
Denote the ground state without / with the twist line as $\ket{\psi}$ and $\ket{\psi}_{Z_2}$. Because for the systems under consideration here, local reduced density matrices of the ground states are all determined by local Hamiltonian terms, we expect that
\begin{enumerate}
\item{Away from the twist line, $\ket{\psi}$ and $\ket{\psi}_{Z_2}$ should look the same.}
\item{Near the twist line, $\ket{\psi}_{Z_2}$ should look like $\prod_{r\in R} \sigma^r_x \ket{\psi}$ where $R$ is a large region with the twist line as part of the border.}
\end{enumerate}
In this way, we can discuss symmetry fluxes in terms of the ground state, instead of the Hamiltonian.

\begin{figure}[htbp]
\begin{center}
\includegraphics[width=8.0cm]{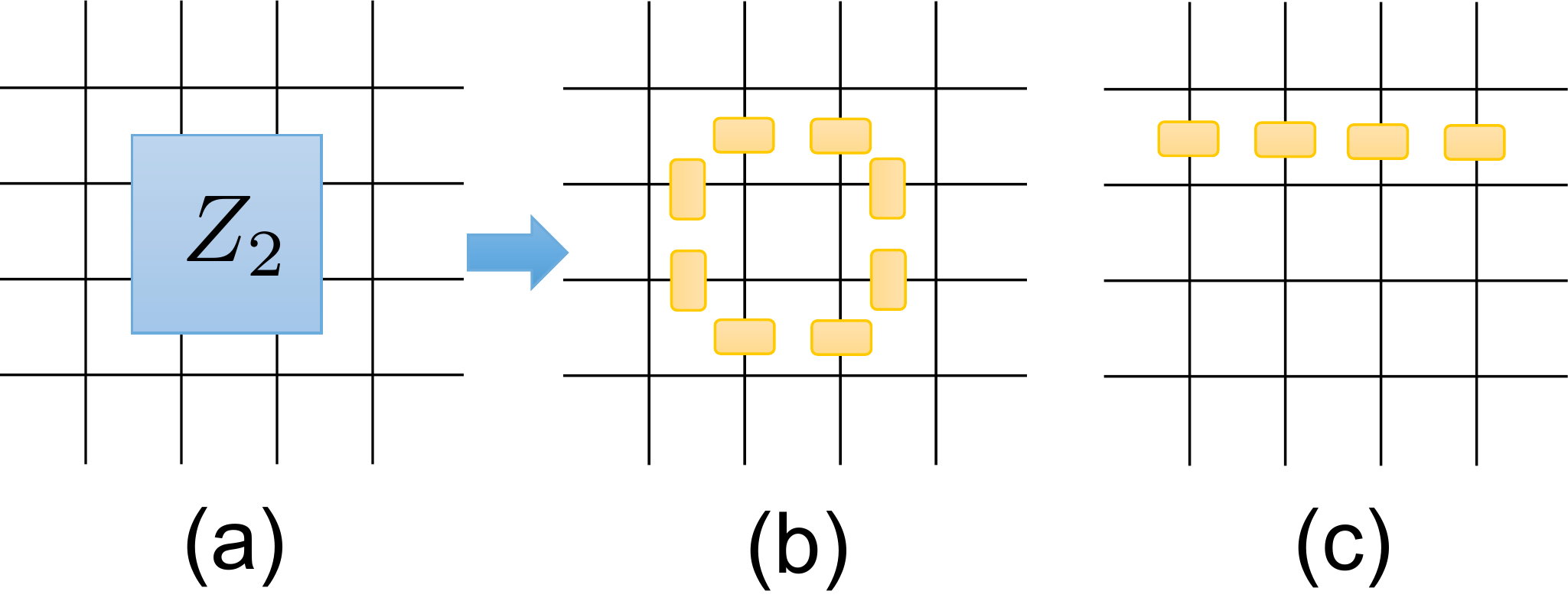}
\caption{(a) Local $Z_2$ symmetry action on a symmetric TPS (b) changes only tensors near the border by the inner symmetry operators; (c) Coupling a TPS to a $Z_2$ symmetry flux corresponds to changing the tensors by inserting the inner symmetry operators along the twist line. For clarity, physical indices are omitted in this figure. All links represent inner indices of the TPS.}
\label{TPS_flux}
\end{center}
\end{figure}  

For simplicity of discussion, we combine every two sites on the A, B sub-lattices and map the system to square lattice structure $\vcenter{\hbox{\includegraphics[scale=0.18]{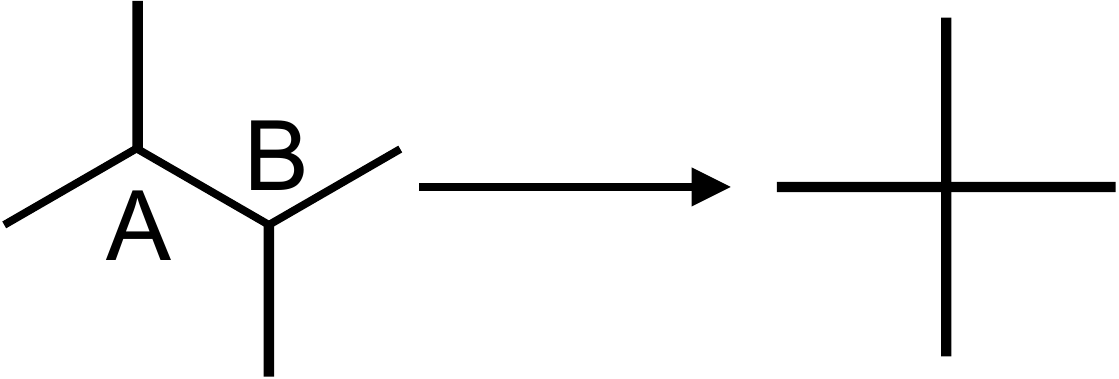}}}$. Each tensor now has four inner indices. The inner symmetry operators are $X\otimes X\alpha$ for the up and left indices (inherited from sub lattice A) and $X\otimes X\bar{\alpha}$ for the down and right indices (inherited from sub lattice B). 

The tensor product representation of the ground state provides a particularly simple way to find $\ket{\psi}_{Z_2}$. As applying symmetry in a region changes the tensors on the border by the inner symmetry operators, as shown in Fig.\ref{TPS_flux} (a) and (b), the tensor product representation of $\ket{\psi}_{Z_2}$ can be obtained from that of $\ket{\psi}$ by inserting the inner symmetry operators along the twist line, as shown in Fig.\ref{TPS_flux} (c). 

\begin{figure}[htbp]
\begin{center}
\includegraphics[width=8.0cm]{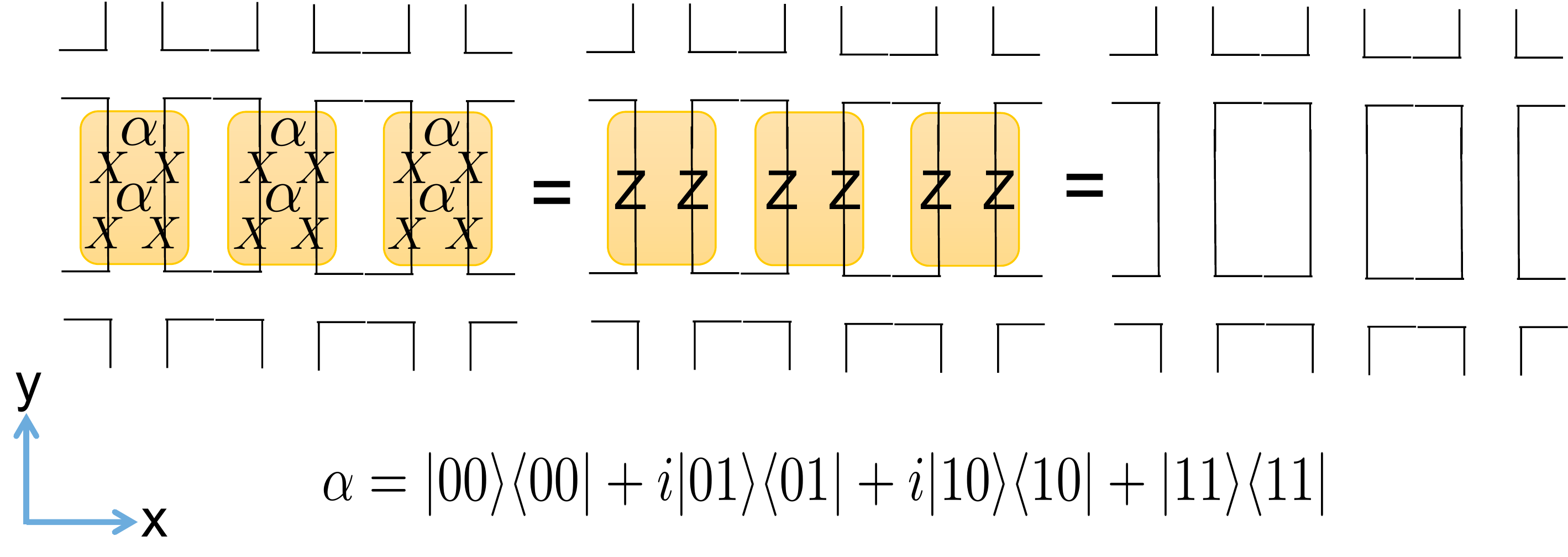}
\caption{In the nontrivial $Z_2$ SPT state, composing two $Z_2$ twist lines in the $x$ direction is equivalent to having no twist lines in the state. The tensor product representation of the state follows from that given in Fig.\ref{fig:Z2_TPS}. Physical indices are omitted in the drawing for clarity.}
\label{fig:Z2f_xx}
\end{center}
\end{figure}

In a square lattice TPS with periodic boundary condition as shown in Fig.\ref{fig:Z2f_xx}, threading a $Z_2$ flux in the $x$ direction (through the nontrivial loop in the $y$ direction) corresponds to inserting inner symmetry operators, $X\otimes X\alpha$ or $X\otimes X\alpha$, along the nontrivial loop in the $x$ direction. Threading a $Z_2$ fluxes in the $y$ direction (through the nontrivial loop in the $x$ direction) corresponds to inserting inner symmetry operators, $X\otimes X\alpha$ or $X\otimes X\bar{\alpha}$, along the twist line of nontrivial loop in the $y$ direction. Composing two twist lines in the same direction should be equivalent to having no flux in this direction. However, as we will see, this equivalence is true only up to a phase factor, which is a topological invariant characterizing the underlying SPT order. A similar procedure of inserting symmetry / gauge twist lines and applying modular transformations to extract topological invariants from the states is discussed in Refs.\onlinecite{Hung2013a,Moradi2014}. 

Consider first a state with $Z_2$ twist lines only in the $x$ direction. Composing two twist lines, we find that the inner symmetry operator on each inner index compose into
\be
\ba {llll}
(X\otimes X\alpha)(X\otimes X\alpha) &= & I  & \text{for trivial $\alpha$} \\
                                                            &= & Z\otimes Z &  \text{for nontrivial $\alpha$}
\ea
\label{Z2Z2}
\ee
With trivial $\alpha$, it is obvious that two $Z_2$ twist lines compose into zero. With nontrivial $\alpha$, which is illustrated in Fig.\ref{fig:Z2f_xx}, the inner symmetry operators on each inner index do not compose into $I$. However, acting $Z\otimes Z$ on all inner indices along a loop does not change the state at all. This is a special inner symmetry of the tensor product representation given in Fig.\ref{fig:Z2_TPS}. Therefore, for both the trivial and nontrivial $Z_2$ SPT states, we see that composing two twist lines in the $x$ direction is equivalent to having no twist line, with no extra phase factor.

\begin{figure}[htbp]
\begin{center}
\includegraphics[width=8.4cm]{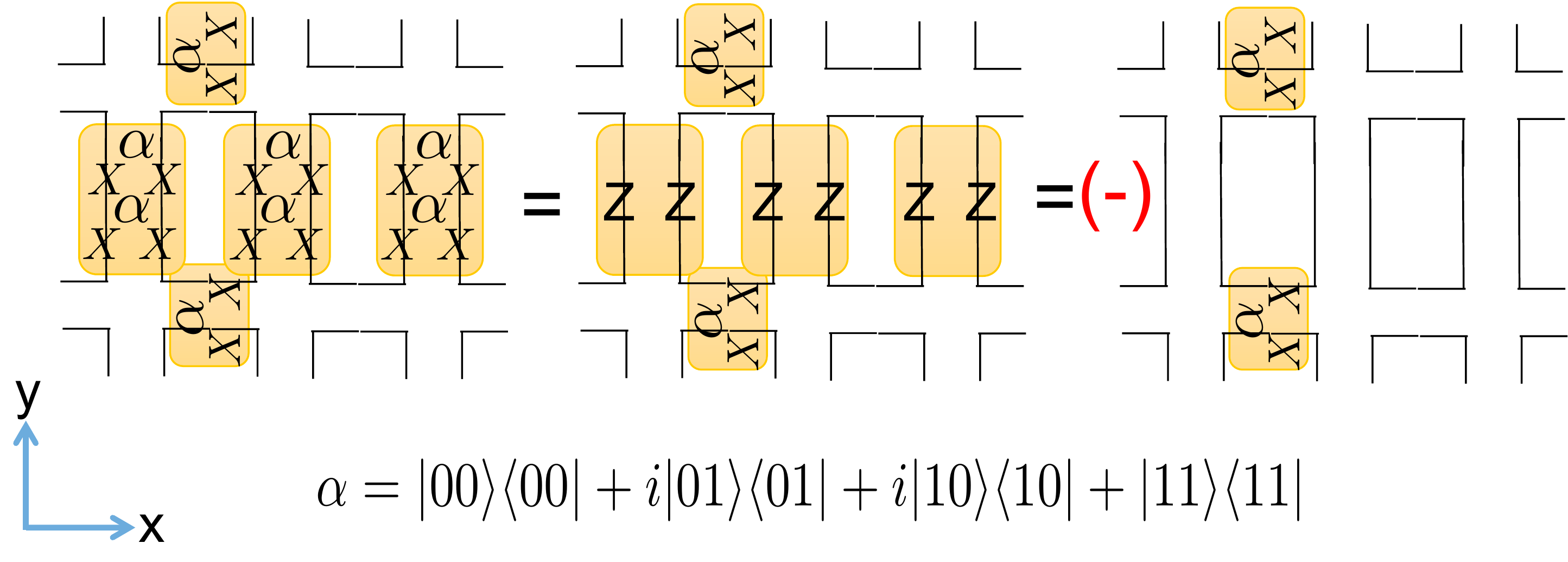}
\caption{In the nontrivial $Z_2$ SPT state, composing two $Z_2$ twist lines in the $x$ direction in the presence of a $Z_2$ twist line in the $y$ direction is equivalent to a state with only a $y$ direction twist line up to a $-1$ phase factor. The tensor product representation of the state follows from that given in Fig.\ref{fig:Z2_TPS}. Physical indices are omitted in the drawing for clarity.}
\label{fig:Z2f_yxx}
\end{center}
\end{figure}

Nontrivial phase factors can arise when we compose two twist lines in the $x$ direction in the presence of a twist line in the $y$ direction. For the trivial SPT state, the phase factor is still $1$ and does not change due to the $y$ direction twist line. However, for the nontrivial SPT state, we can see from Fig.\ref{fig:Z2f_yxx} that a $-1$ phase factor arises. The inner symmetry operators along the $x$ direction compose into $Z\otimes Z$ on each inner index.  While the $Z\otimes Z$ operators in the middle of the loop keeps the tensor product state invariant, the two on the two sides of the $y$ direction twist line results in a $-1$ phase factor due to the presence of the $X\otimes X \alpha$ operator in the $y$ direction. Therefore, for the nontrivial SPT state, the composition of twist lines in the $x$ direction is projective (with a $-1$ phase factor) in the presence of a twist line in the $y$ direction. 

We can interpret this $\pm 1$ phase factor as related to the $Z_2^2$ symmetry transformation on each $Z_2$ symmetry defect locally which exists at the end of $Z_2$ twist lines. For the $Z_2$ SPT state, we know that if we create $Z_2$ symmetry defects in the bulk and promote them into deconfined excitations by gauging the symmetry, they have bosonic or semionic exchange statistics in the trivial and nontrivial SPT state. Equivalently, we can say that each $Z_2$ symmetry defect carries $0$ or $1/2$ $Z_2$ charge in these two phases. That is, each $Z_2$ symmetry defect gets a $1$ or $i$ phase factor under the local action of $Z_2$ symmetry. If we apply the $Z_2$ symmetry locally twice on one $Z_2$ symmetry defect, we expect to get a phase factor of $1$ and $-1$ respectively, which is exactly what we obtained from the twist line composition process discussed above. That is, the projective phase factor coming from the composition of two $Z_2$ twist lines in the $x$ direction in the presence of a $Z_2$ twist line in the $y$ direction can be interpreted as the $Z_2^2$ transformation on each $Z_2$ symmetry defect locally. 

To understand this connection, we can imagine cutting the system open along the nontrivial loop in the $y$ direction and turn the torus into a cylinder. Due to the existence of the $Z_2$ twist line in the $y$ direction, $Z_2$ symmetry defects are present at the top and bottom end of the cylinder. Now inserting two $Z_2$ twist lines in the $x$ direction at the bottom end of the cylinder as show in Fig.\ref{fig:Z2f_yxx} is effectively measuring the $Z_2^2$ quantum number of one $Z_2$ symmetry defect. Note that if we actually apply $Z_2$ symmetry $\prod \sigma_x$ twice in a region near the bottom end, we will not be able to see the nontrivial phase factor because $(\prod \sigma_x)^2$ is always equal to $1$. Similar situations happen with time reversal symmetry in the examples discussed in later sections. 

Similar to the 1D case, this result applies not only to the fixed point tensors shown in this section, but also to those away from the fixed point. The topological properties of the inner symmetry operators remain the same when the tensors are perturbed away from the fixed point form.

Finally, we want to comment that there are two differences between the 1D time reversal SPT example and the 2D $Z_2$ SPT example: 1. in composing time reversal twists we need to take complex conjugation on one of the twist while composing unitary $Z_2$ twists does not involve such a step; 2. to see the nontrivial phase factor, for the 1D SPT state we are simply composing two twists while for the 2D case we are composing twists in the presence of another twist in the orthogonal direction and thus investigating the relation between them. In our following study of 2D phases with time reversal symmetry, these two features need to be combined.

\subsection{Local action and twists of time reversal in 2D}
\label{2DT}

Now we are ready to generalize this procedure to time reversal symmetry in 2D. In 2D, there is only a trivial time reversal symmetric SPT phase\cite{Chen2012a,Chen2013a,Lu2012} and we use it as an example to illustrate the basic ideas. In section \ref{2DSPT_Z2Z2T} and \ref{2DSET_Z2T}, we discuss the more interesting cases of nontrivial SPT phases with $Z_2\times Z_2^T$ symmetry and $Z_2$ gauge theories with time reversal symmetry.

Consider a 2D square lattice with four spin $1/2$'s per site and the four spin $1/2$'s at the corners of each plaquette form an entangled state $\ket{0000}+\ket{1111}$. The total wave function is hence
\be
\ket{\psi}=\prod_{\square} \left(\ket{0000}+\ket{1111}\right) \label{Z2Twf}
\ee
where the product is over all plaquettes $\square$.
Time reversal symmetry acts by first taking complex conjugation in the $\ket{0}$, $\ket{1}$ basis, and then applying $\sigma_x$ to each spin. The state is obviously short range entangled and time reversal invariant. We can think of the four spins around a plaquette as a time reversal domain. The wave function is hence an equal weight superposition of all domain configurations. Of course, there are simpler states with the same SPT order, but here we use this form of short range entangled wave function because this is the standard form of SPT wave function and all SPT states can be written in a similar way. (The $Z_2$ SPT state in Eq.\ref{Z2wf0} and \ref{Z2wf1} can also be put into this form with a local basis transformation.)

\begin{figure}[htbp]
\begin{center}
\includegraphics[width=4cm]{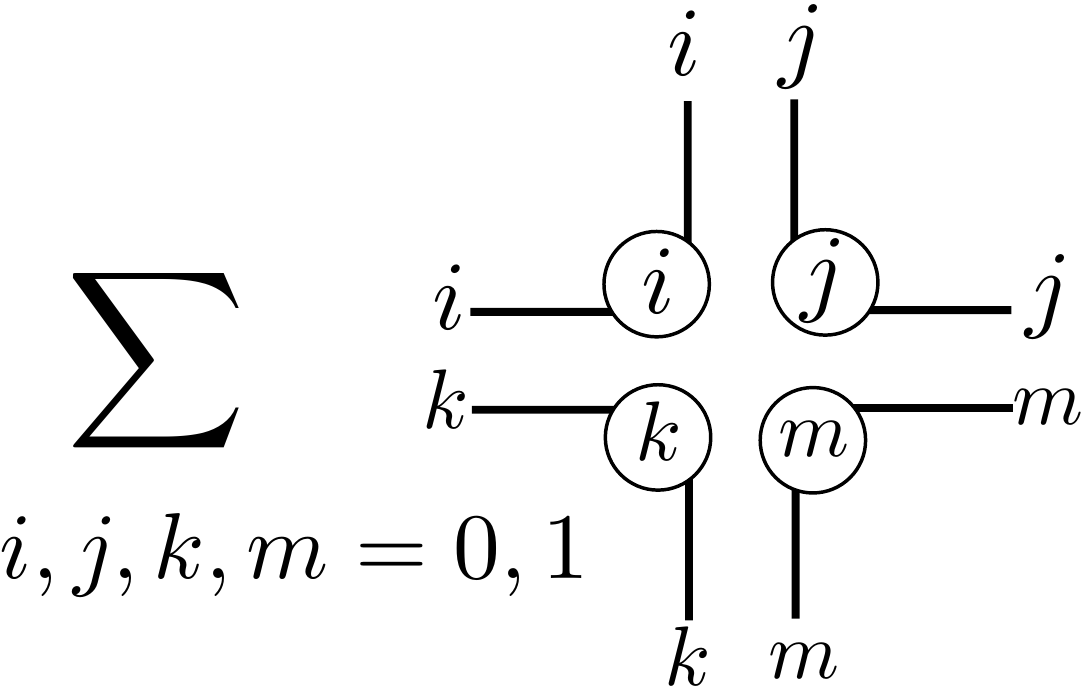}
\caption{Tensor representing 2D trivial SPT state with time reversal symmetry. The labels in circles are physical indices and the labels at the end of the links are inner indices.}
\label{fig:Z2T_TPS}
\end{center}
\end{figure}

To define local time reversal symmetry action, we need the tensor product representation of the state, which gives us a way to divide the coefficient of the wave function into local pieces. The tensors at each site can be chosen as shown in Fig.\ref{fig:Z2T_TPS}. 

\begin{figure}[htbp]
\begin{center}
\includegraphics[width=4cm]{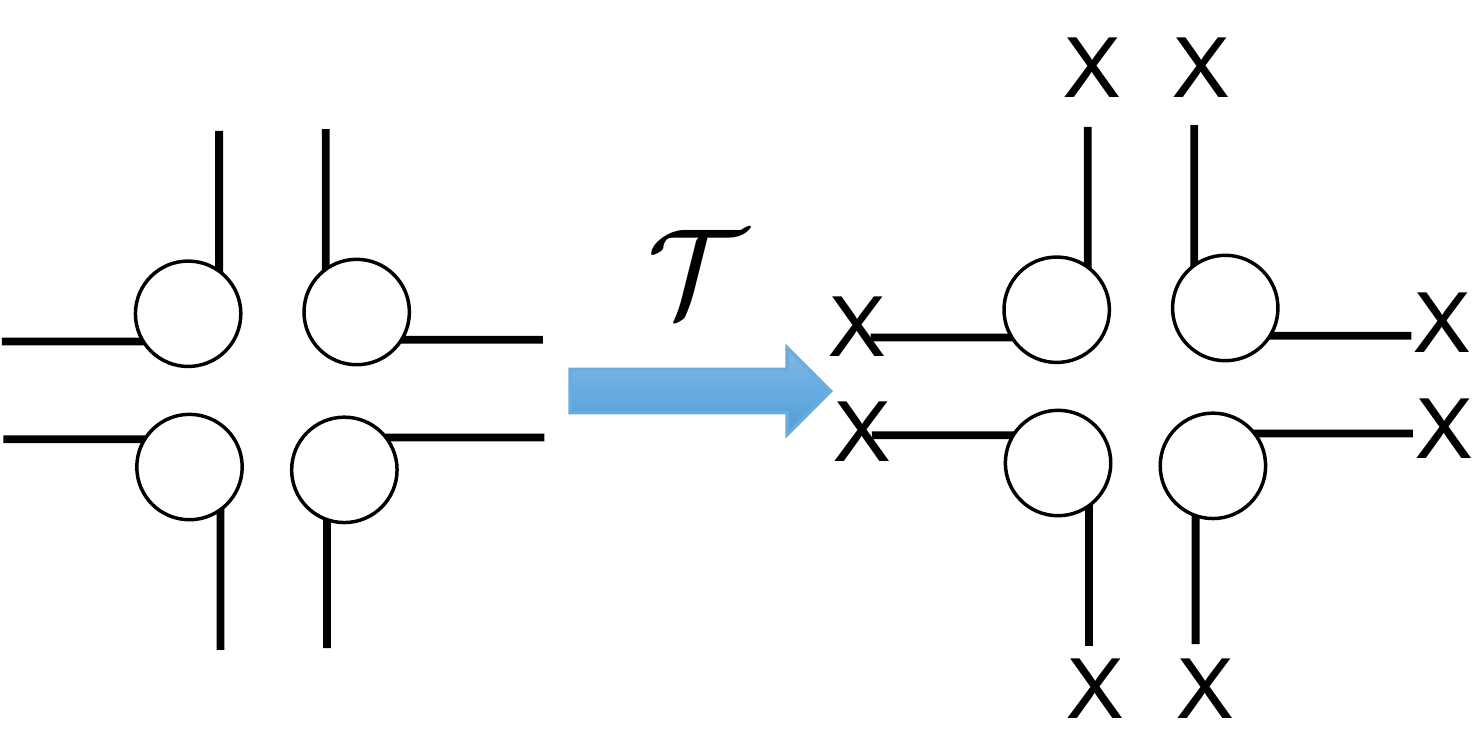}
\caption{Acting time reversal on one site induces `inner symmetry operators' $X\otimes X$ on the inner indices.}
\label{fig:sym_Z2TTPS}
\end{center}
\end{figure}

With the tensor product representation, we can now define local action of time reversal symmetry as: 1. taking complex conjugation of the tensors  2. acting $\sigma_x K$ on the physical basis. The induced `inner symmetry operators' on the tensors are $X\otimes X$ on each inner index, as shown in Fig.\ref{fig:sym_Z2TTPS}. From this relation, we can see that the state is invariant under global time reversal symmetry action, because the inner symmetry operators cancel with each other if time reversal is applied globally. Moreover, acting time reversal locally in a subregion changes only the tensors along the border.

Knowing how time reversal acts locally on the state, we can insert twist lines and couple the state to time reversal fluxes through the nontrivial loops in the system. Similar to the $Z_2$ case discussed previously, we expect that
\begin{enumerate}
\item{Away from the twist line, the states with and without time reversal flux should look the same.}
\item{Near the twist line, the state with time reversal flux should look like the state without flux but with time reversal symmetry acting locally in region $R$, where $R$ is a large region with the twist line as part of the border.}
\end{enumerate}

For the time reversal symmetric state discussed above, this can be realized by inserting inner symmetry operators $X\otimes X$ into the inner indices along a nontrivial loop (in the $x$ or $y$ direction). We expect the projective composition rule of time reversal twist lines to reflect the universal topological properties of the state. Note that because we are considering time reversal twists here, when composing two inner symmetry operators, we should take complex conjugation of the first one. It is easy to see that two copies of the inner symmetry operators $X\otimes X$ naturally compose into identity, therefore time reversal twist lines in this state do not have nontrivial projective composition, as we would expect for a state with trivial SPT order.

We want to make a comment about how the gauge choice for the tensor product representation affects the result. For unitary symmetries discussed in the previous section, changing the gauge of the tensors do not affect the projective composition rule at all. Suppose that we change the gauge of the tensors by an invertible matrix $N$. The inner symmetry operators all change by conjugation with $N$ and $N^{-1}$ and their composition and commutation relations remain the same. 

For the time reversal symmetry discussed in this section, changing the gauge of the tensor leaves the result almost invariant, except for one subtlety: there are certain gauge choices of the tensors such that local time reversal symmetry action results in a null state. Suppose that we change the gauge of the upper index of the tensor in Fig.\ref{fig:Z2T_TPS} by $e^{i\frac{\pi}{4}X} \otimes I$. Then applying time reversal symmetry locally on one site leads to inner symmetry operators $iI\otimes X$ on the upper index and $X\otimes X$ on the other indices. When we try to connect the tensor on different sites and find the total wave function after local time reversal symmetry action, we find that the resulting state has zero amplitude. Therefore, we need to exclude these possibilities and require that local time reversal symmetry action always results in a nonzero state.

As long as this nonzero condition is satisfied, gauge change of the tensors does not affect the composition of time reversal twist lines. For the time reversal symmetric trivial SPT state discussed above, changing the gauge by an invertible matrix $N$ changes the inner symmetry operators to $N^*(X\otimes X)N^{-1}$. Two such operators composed together is still the identity 
\be
(N^*(X\otimes X)N^{-1})^* N^*(X\otimes X)N^{-1} = I
\ee
Therefore, our discussion is independent of the gauge choice of the tensor product representation (as along as local time reversal symmetry action results in a nonzero state).

\subsection{Example: 2D SPT with $Z_2\times Z_2^T$ symmetry} 
\label{2DSPT_Z2Z2T}

Now let's study a more interesting example: a 2D SPT state with $Z_2\times Z_2^T$(time reversal) symmetry. The 2D SPT phases with $Z_2\times Z_2^T$ symmetry have a $\mathbb{Z}_2\times \mathbb{Z}_2$ classification.\cite{Chen2012a,Chen2013a,Lu2012} The trivial and nontrivial SPT order with pure $Z_2$ symmetry accounts for the first $\mathbb{Z}_2$ in the classification. The second $\mathbb{Z}_2$ in the classification corresponds to the $Z_2$ symmetry defects in the bulk of the state transforming as $\cT^2 = \pm 1$ under time reversal.\cite{Chen2014} (This kind of `local Kramer degeneracy' has also been studied in Ref.\onlinecite{Levin2012a}.) To distinguish states with trivial and nontrivial SPT order under the $Z_2$ part of the symmetry, we can insert $Z_2$ twist lines along the nontrivial loops and study their projective composition rules as discussed in section \ref{Z2}. In this section, we are going to show how to determine whether $\cT^2 = 1$ or $-1$ on each $Z_2$ symmetry defect from the projective composition rule of the time reversal twist lines in the presence of a unitary $Z_2$ twist line.

\begin{figure}[htbp]
\begin{center}
\includegraphics[width=8cm]{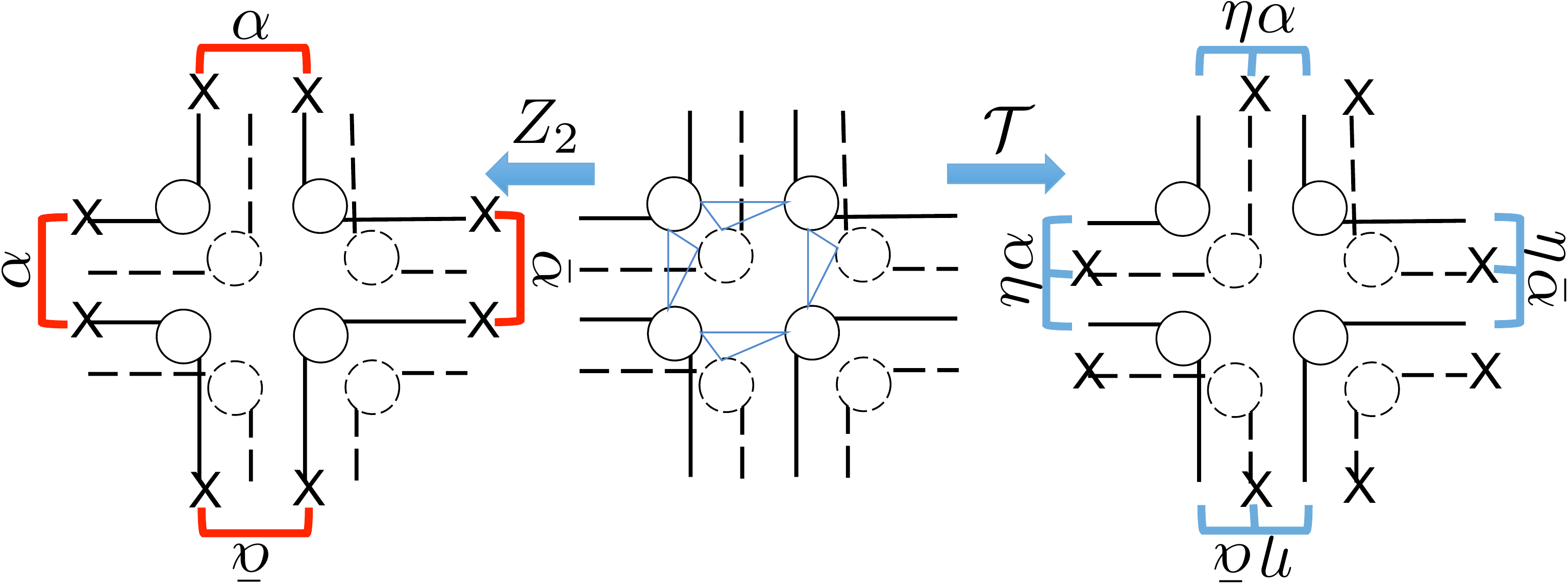}
\caption{Local symmetry action on the tensor representing the SPT state with $Z_2\times Z_2^T$ symmetry. The solid ($Z_2$) part of the tensor follows from that given in section \ref{Z2} and the dashed (time reversal) part of the tensor is the same as that in Fig.\ref{fig:Z2T_TPS}. Local $Z_2$ and time reversal symmetry induce changes to the tensors as shown on the left and right hand side of this figure.}
\label{fig:sym_Z2Z2T}
\end{center}
\end{figure}

Consider a $Z_2\times Z_2^T$ SPT state on the square lattice with two sets of spin $1/2$ degrees of freedom $\sigma$ (solid circles in Fig.\ref{fig:sym_Z2Z2T}, which we call the $Z_2$ spins) and $\tau$ (dashed circles in Fig.\ref{fig:sym_Z2Z2T}, which we call the time reversal spins).  The wave function is a product of two parts: the $Z_2$ part $\ket{\psi_{Z_2}}$ and the time reversal part $\ket{\psi_T}$
\be
\ket{\psi} = \ket{\psi_{Z_2}} \otimes \ket{\psi_T}
\ee
The $Z_2$ part formed by the $\sigma$ spins takes the same form as discussed in section \ref{Z2}, which can have either trivial or nontrivial $Z_2$ SPT order. $Z_2$ symmetry acts as $\sigma_x$ on each $Z_2$ spin and does not affect the time reversal part of the wave function. The local $Z_2$ symmetry action, the $Z_2$ symmetry twist line and their projective composition rules follow directly from section \ref{Z2}, for both the trivial and nontrivial $Z_2$ SPT order.
The time reversal part $\ket{\psi_T}$ of the wave function is formed by the $\tau$ spins and is an equal weight superposition of time reversal domain configurations, as discussed in section \ref{2DT}. The tensor product representation of the time reversal part is hence the same as that given in Fig.\ref{fig:Z2T_TPS}. The tensor product representation of the total wave function is a product of the $Z_2$ part and the time reversal part. While the $Z_2$ part and the time reversal part are decoupled in the wave function, they can be intertwined in the definition of time reversal symmetry action, hence giving rise to possibly nontrivial transformation of $Z_2$ symmetry defects under time reversal. We will see how this works in detail below.

The SPT order related to the second $\mathbb{Z}_2$ in the classification ($\cT^2=\pm 1$ on $Z_2$ symmetry defects) is encoded in the way time reversal symmetry acts on the wave function.
The global time reversal symmetry can act on $\sigma$ and $\tau$ spins together. It is composed of three parts: 1. taking complex conjugation in the $\ket{0}$, $\ket{1}$ basis of $\sigma$ and $\tau$ spins for the whole wave function 2. applying $\tau_x$ to all $\tau$ spins 3. applying phase factors $\eta$ in the $\ket{0}$, $\ket{1}$ basis to the three spins connected by each triangle in Fig.\ref{fig:sym_Z2Z2T}. $\eta$ involves two $\sigma$ spins and one $\tau$ spin and has two possibilities: 
\be
\ba{l}
\eta=1\ \text{for all states, or,} \\
\eta=-1 \ \text{on} \ \ket{110} \ \text{and} \ \ket{011}, \ \eta=1 \ \text{otherwise}
\ea
\label{eta}
\ee
The ordering of spins in the definition of $\eta$ is $\sigma\tau\sigma$. $\eta$ is symmetric with respect to the exchange of the two $\sigma$'s. 
If $\eta=1$ for all states, then the $Z_2$ part and the time reversal part of the wave function are decoupled in the time reversal symmetry action. Therefore, the $Z_2$ symmetry defect has to transform trivially as $\cT^2 =1$ under time reversal and hence the state has trivial SPT order in the second $\mathbb{Z}_2$ classification. With the nontrivial $\eta$, the $Z_2$ symmetry defect transforms as $\cT^2 =-1$ and the state has nontrivial SPT order in the second $\mathbb{Z}_2$ classification.

To understand how the nontrivial $\eta$ is related to the $\cT^2=-1$ transformation of the $Z_2$ symmetry defects, we can think of the 
time reversal $\tau$ spin involved in each triangle as living between the $Z_2$ domains formed by the two $\sigma$ spins in the same triangle
as shown in Fig.\ref{fig:sym_Z2Z2T}. From the definition of $\eta$ we see that, if $\tau$ is on a $Z_2$ domain wall, then time reversal acts on it as $i\tau_y K$ which squares to $-1$. If $\tau$ is not on a $Z_2$ domain wall, then time reversal acts on it as $\tau_x K$ which squares to $1$. Along the domain wall, the $\cT^2=-1$ $\tau$ spins form time reversal singlets and the total wave function is a superposition of all $Z_2$ configurations with domain walls decorated by the time reversal singlets. When $Z_2$ symmetry defects are inserted in the bulk of the system, the $Z_2$ domain wall ends, leaving un-paired $\cT^2=-1$ $\tau$ spins at each of the defect point. Therefore, with nontrivial $\eta$, each $Z_2$ symmetry defect in the bulk transforms as $\cT^2=-1$ under time reversal symmetry.

This feature can be identified from the projective composition rule of the time reversal twist lines in the state. First, let us find the local time reversal symmetry action on the representing tensors of the state. The local time reversal symmetry action on the tensors is given by taking complex conjugation on the tensor and applying $\tau_x$, $\eta$ and complex conjugation on the physical basis states. Fig.\ref{fig:sym_Z2Z2T} illustrates the inner symmetry operators induced by the local time reversal symmetry action on the tensor, which reads
\be
M_{\cT} = (X^{\tau} \otimes X^{\tau}) \cdot \eta^{\sigma \tau \sigma} \cdot \alpha^{\sigma\sigma} (\text{or} \ \bar{\alpha}^{\sigma\sigma})
\ee
Here, the $\sigma$ and $\tau$ superscripts label the inner indices in the $Z_2$ and time reversal part of the tensor. The $\alpha^{\sigma\sigma}(\text{or} \ \bar{\alpha}^{\sigma\sigma})$ part comes from taking complex conjugation of the $Z_2$ part of the tensor.
On the left hand side of the figure, we also illustrate the inner symmetry operator induced by local $Z_2$ symmetry action, which is
\be
M_{Z_2}=(X^{\sigma}\otimes X^{\sigma}) \cdot \alpha^{\sigma\sigma} (\text{or} \ \bar{\alpha}^{\sigma\sigma})
\ee
as discussed in section \ref{Z2}.

Suppose that we compose two time reversal twist lines along the $x$ direction. The composition of two time reversal inner symmetry operators gives
\be
\ba{ll}
 & M_{\cT}M^*_{\cT} \\
= & [( X^{\tau} \otimes X^{\tau}) \cdot \eta^{\sigma \tau \sigma} \cdot \alpha^{\sigma\sigma}] [( X^{\tau} \otimes X^{\tau}) \cdot \eta^{\sigma \tau \sigma} \cdot \alpha^{\sigma\sigma}] ^*  \\
= & I \ \ \text{for trivial $\eta$} \\
= & Z^{\sigma} \otimes Z^{\sigma} \ \ \text{for nontrivial $\eta$}
\ea
\ee
The result is similar to the composition of two $Z_2$ twist lines discussed in Eq.\ref{Z2Z2}.
Therefore, if the system has only two time reversal twist lines in the $x$ direction, then their composition is equivalent to having no twist line. 

\begin{figure}[htbp]
\begin{center}
\includegraphics[width=8.4cm]{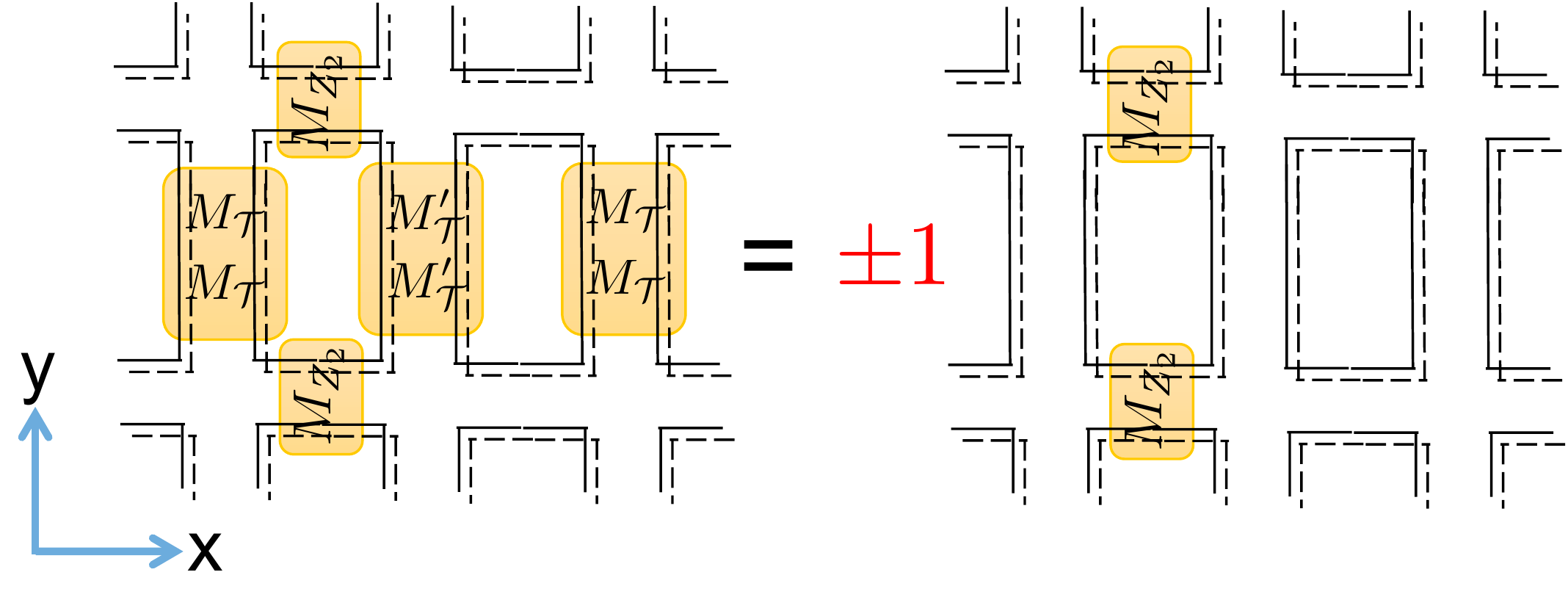}
\caption{In the $Z_2\times Z_2^T$ SPT states, composing two time reversal twist lines in the $x$ direction in the presence of a $Z_2$ twist line in the $y$ direction is equivalent to a state with only a $y$ direction $Z_2$ twist line up to a $\pm1$ phase factor, which corresponds to the $\cT^2=\pm 1$ transformation law on each $Z_2$ symmetry defect. The tensor product representation of the state follows from that given in Fig.\ref{fig:sym_Z2Z2T}. Physical indices are omitted in the drawing for clarity.}
\label{fig:Z2Z2Tf_yxx}
\end{center}
\end{figure}

\begin{figure}[htbp]
\begin{center}
\includegraphics[width=8cm]{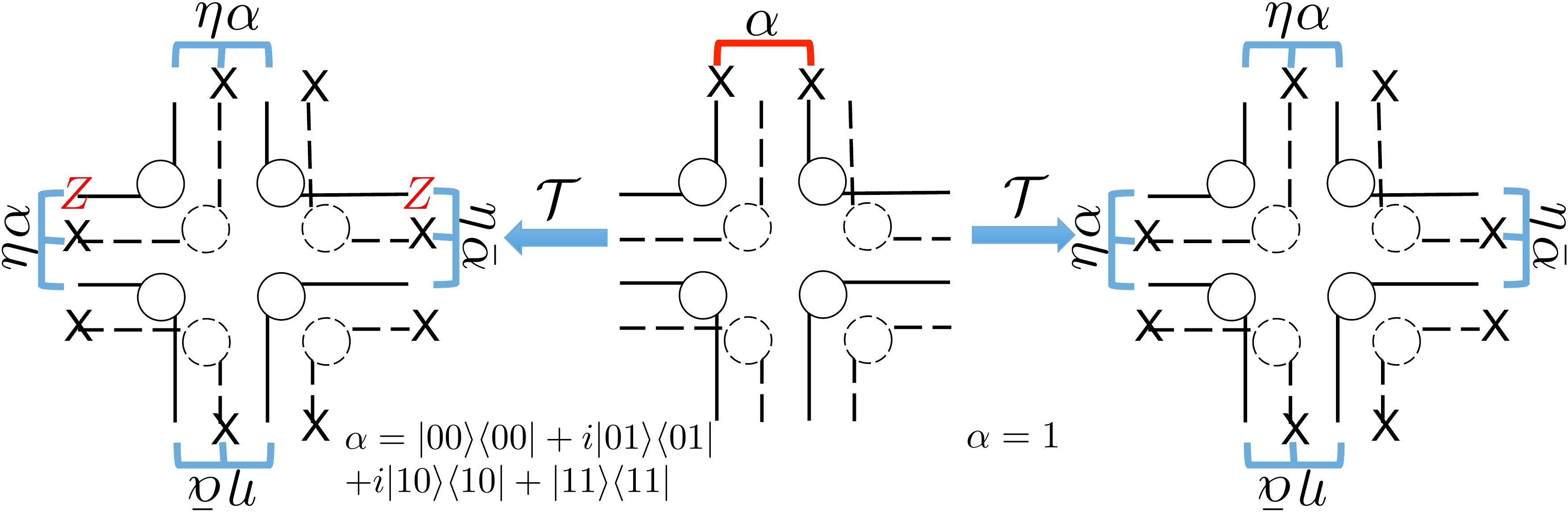}
\caption{Local time reversal action on the tensor representing $Z_2\times Z_2^T$ SPT state with $Z_2$ twist line inserted on the top side. When $\alpha$ is trivial (right hand side), the induced inner symmetry operator remains the same as in Fig.\ref{fig:sym_Z2Z2T}; when $\alpha$ is nontrivial (left hand side), the induced inner symmetry operator changes by $Z^{\sigma}\otimes I^{\sigma}$.}
\label{fig:sym_TZ2f}
\end{center}
\end{figure}

However, if we are composing two time reversal twist lines in the $x$ direction in the presence of a $Z_2$ symmetry twist line in the $y$ direction, a $-1$ phase factor arises for nontrivial $\eta$. From Fig.\ref{fig:Z2Z2Tf_yxx} it is easy to see that, when $M_{\cT}M^*_{\cT}=I$ with trivial $\eta$, the composition results in a $+1$ phase factor while when $M_{\cT}M^*_{\cT}=Z^{\sigma} \otimes Z^{\sigma}$ with nontrivial $\eta$, the composition results in a $-1$ phase factor. There is a small complication which we need to explain: the time reversal inner symmetry operator may take a different form ($M'_{\cT}$) at the crossing point with the $Z_2$ twist line than everywhere else on the time reversal twist line ($M_{\cT}$). This is because with the inserted $Z_2$ twist line, local time reversal symmetry action may result in different inner symmetry operators. In particular, as shown in Fig.\ref{fig:sym_TZ2f}, when $\alpha$ is trivial, $M'_{\cT}=M_{\cT}$; when $\alpha$ is nontrivial, $M'_{\cT}=M_{\cT}(Z^{\sigma}\otimes I^{\sigma})$. Explicit calculation shows that our result regarding the composition of time reversal twist lines is still valid.

In section \ref{Z2}, we interpreted the projective phase factor in the composition of two $Z_2$ twist lines in the presence of another $Z_2$ twist line as the $Z_2^2$ value of a $Z_2$ symmetry defect. Similarly, here we interpret the projective phase factor in the composition of two time reversal twist lines in the presence of another $Z_2$ twist line as the $\cT^2$ value of a $Z_2$ symmetry defect.

Note that the projective phase factor in the composition of two time reversal twist lines depends only on $\eta$ and not on $\alpha$. This is consistent with the $\mathbb{Z}_2\times \mathbb{Z}_2$ structure of the $Z_2\times Z_2^T$ SPT classification. $\alpha$ is responsible for the projective phase factor in the composition of two $Z_2$ twist lines which is related to the first $\mathbb{Z}_2$ classification while $\eta$ is related to the second. 

Finally we want to comment that this result is independent of the gauge choice of the representing tensors, as long as local time reversal action leads to a nonzero state. This is similar to the cases discussed in the previous section.

\subsection{Example: Toric code and double semion topological order with time reversal symmetry} 
\label{2DSET_Z2T}

In the previous sections, we studied symmetry protected topological phases with short range entanglement. In this section, we study an example of long range entangled state and demonstrate that the notions of local time reversal action and time reversal twist lines still work. In particular, we are going to gauge the $Z_2$ symmetry in the examples discussed in the last section and study $Z_2$ gauge theories with time reversal symmetry. These are examples of the so-called 'symmetry enriched topological' (SET) orders.\cite{Essin2013, Lu2013} 

In the above discussion, we have seen that there are four ($\mathbb{Z}_2\times \mathbb{Z}_2$) SPT phases in 2D with $Z_2\times Z_2^T$ symmetry. Note that even with just $Z_2$ symmetry, there are two phases. In that case, gauging the $Z_2$ symmetry leads to two distinct topological orders, the topological order of the usual $Z_2$ gauge theory ($Z_2$ topological order like in the toric code state\cite{Kitaev2003}) and the twisted $Z_2$ gauge theory (double semion topological order)\cite{Levin2012}.  Now, in the presence of time reversal symmetry, the SPT classification implies that there could be an additional two fold distinction. For the usual $Z_2$ topological order, this is readily interpreted as arising when each bosonic $Z_2$ gauge defect (gauge flux excitation) transforms either linearly with $\cT^2=1$ or projectively with $\cT^2=-1$ under time reversal symmetry. For the twisted $Z_2$ gauge theory, such a distinction breaks down because the $Z_2$ gauge defects have semionic statistics and are not time reversal invariant. Time reversal acting on the semions turns them into anti-semions. Therefore the $\cT^2$ value on the gauge defects are no longer well defined and there is only one SET phase with twisted $Z_2$ gauge theory and time reversal symmetry.\cite{Burnell2015} We will see below how this distinction between the two types of $Z_2$ gauge theories can be extracted by studying the time reversal symmetry twist lines in the ground state.

\begin{figure}[htbp]
\begin{center}
\includegraphics[width=8cm]{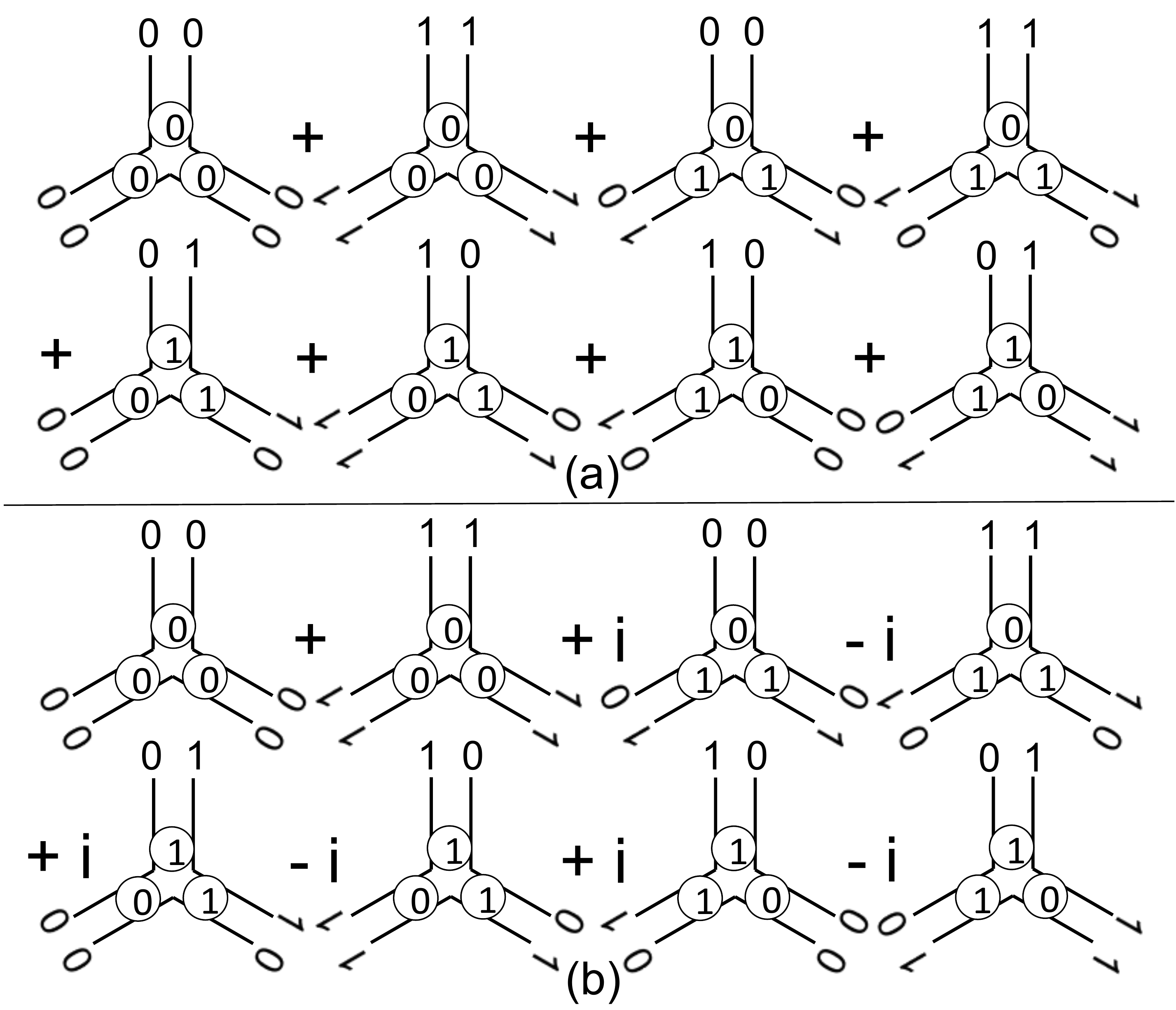}
\caption{The tensor product representation of (a) the toric code and (b) the double semion $Z_2$ gauge theory. The labels in circles are physical indices and the labels at the end of the links are inner indices.}
\label{fig:Z2g_TPS}
\end{center}
\end{figure}

A duality transformation from domain degrees of freedom to domain wall degrees of freedom maps a $Z_2$ symmetric state to a $Z_2$ gauge theory.\cite{Levin2012} The tensor product representation of the toric code $Z_2$ gauge theory\cite{Kitaev2003}  obtained from this mapping is given in Fig.\ref{fig:Z2g_TPS} (a) while that of the double semion $Z_2$ gauge theory is given in Fig.\ref{fig:Z2g_TPS} (b). The tensor for the toric code state can be obtained from that for the double semion state by dropping all the $i$ and $-i$ phase factors. For simplicity of discussion, we again combine every pair of $A$, $B$ lattice sites into one $\vcenter{\hbox{\includegraphics[scale=0.18]{hex2sqr}}}$ and map the system to square lattice.

\begin{figure}[htbp]
\begin{center}
\includegraphics[width=4cm]{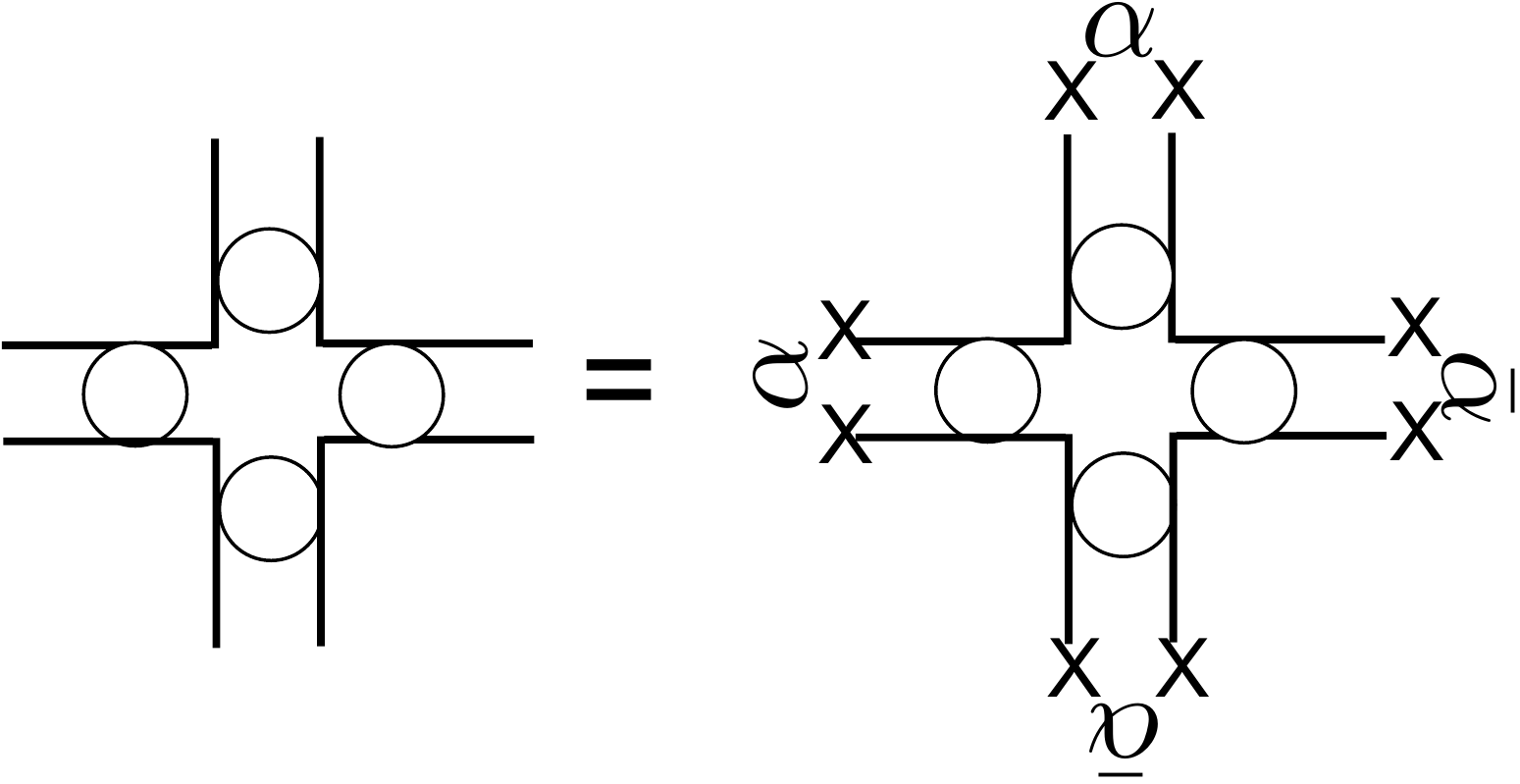}
\caption{The tensors representing the $Z_2$ gauge theories are invariant under inner symmetry operators $(X\otimes X)\alpha$ and $(X\otimes X)\bar{\alpha}$.}
\label{fig:sym_Z2g}
\end{center}
\end{figure}

While for the $Z_2$ SPT states, the representing tensors change by the inner symmetry operators $(X\otimes X)\alpha$ and $(X\otimes X)\bar{\alpha}$ under local $Z_2$ action, the tensors for the $Z_2$ gauge theories are invariant under such inner symmetry operators, as shown in Fig.\ref{fig:sym_Z2g}.

\begin{figure}[htbp]
\begin{center}
\includegraphics[width=8cm]{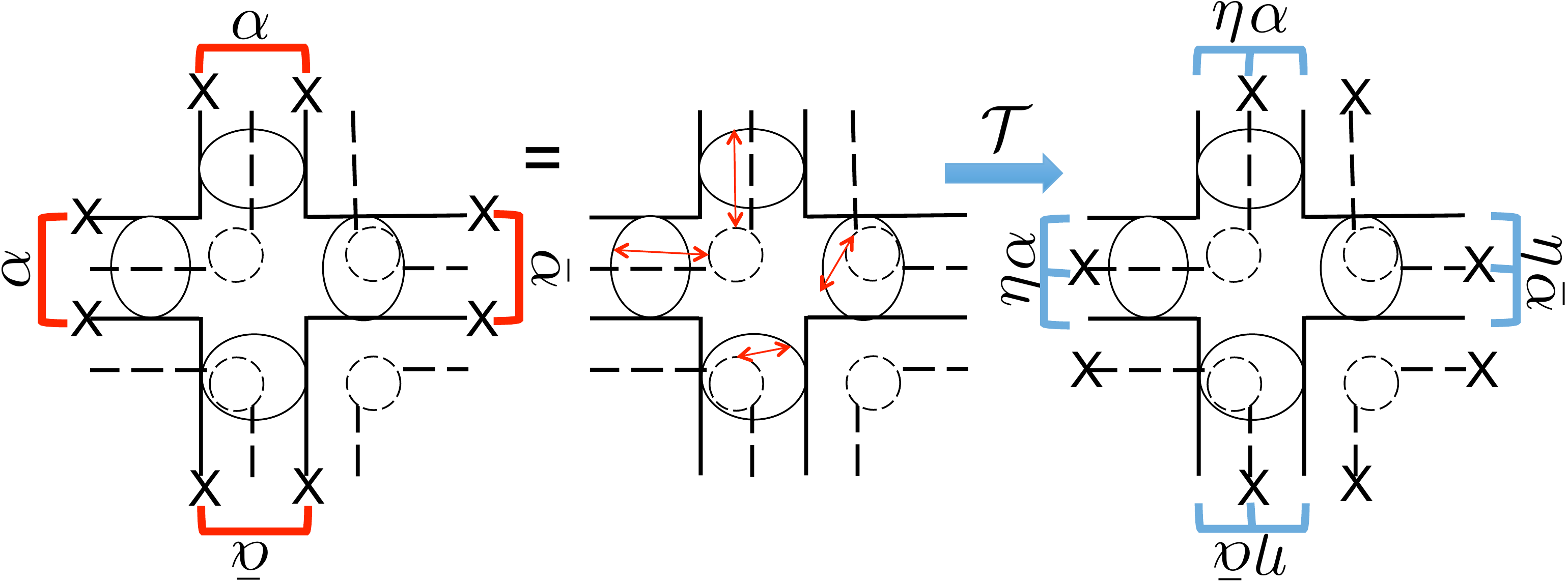}
\caption{Local symmetry action on the tensor representing the $Z_2$ gauge theory with time reversal symmetry. The solid ($Z_2$) part of the tensor follows from that in Fig.\ref{fig:Z2g_TPS} and the dashed (time reversal) part of the tensor is the same as in Fig.\ref{fig:Z2T_TPS}. Local time reversal symmetry induces changes to the tensors as shown on the right hand side of this figure. The left hand side shows an inner invariance of the tensor related to the $Z_2$ gauge symmetry of the state.}
\label{fig:sym_Z2gT}
\end{center}
\end{figure}

To write down a $Z_2$ gauge theory with extra time reversal structure, we introduce time reversal spins $\tau$ into the state such that the tensors representing the state is composed of the $Z_2$ part given above and the time reversal part given in Fig.\ref{fig:Z2T_TPS}. Global time reversal symmetry action contains three parts: 1. taking complex conjugation in the $\ket{0}$, $\ket{1}$ basis of the $Z_2$ and time reversal spins 2. acting $\tau_x$ on all the time reversal spins 3. applying phase factors $\eta'$ in the $\ket{0}$, $\ket{1}$ basis to each pair of $Z_2$ and time reversal spin connected by red arrows in Fig.\ref{fig:sym_Z2gT}.
$\eta'$ involves one $Z_2$ spin and one time reversal spin and has two possibilities:
\be
\ba{l}
\eta'=1\ \text{for all states or,} \\
\eta'=-1 \ \text{on} \ \ket{11}, \ \eta'=1 \ \text{otherwise}
\ea
\label{eta'}
\ee
Note that $\eta'$ can be obtained from $\eta$ in Eq.\ref{eta} by changing the $Z_2$ domain degrees of freedom to the $Z_2$ domain wall degrees of freedom. 

Local time reversal symmetry action then takes complex conjugation of the tensors and applies $\tau_x$, $\eta'$ and complex conjugation to the physical degrees of freedom. The transformation of the tensor under local time reversal action is shown in the right part of Fig.\ref{fig:sym_Z2gT}. The induced inner symmetry operator is the same as that in the $Z_2\times Z_2^T$ SPT state. The left part of Fig.\ref{fig:sym_Z2gT} shows an inner invariance of the tensor which is related to the $Z_2$ gauge symmetry of the state. 

Now we can create time reversal twist lines by inserting 
\be
M_{\cT} = (X^{\tau}\otimes X^{\tau})\cdot \eta^{\sigma\tau\sigma}\cdot \alpha^{\sigma\sigma} (\text{or} \ \bar{\alpha}^{\sigma\sigma})
\ee
into the inner indices along a nontrivial loop. Moreover, we can create $Z_2$ gauge twist lines by inserting 
\be
M_{Z_2}=(X^{\sigma}\otimes X^{\sigma})\cdot \alpha^{\sigma\sigma} (\text{or} \ \bar{\alpha}^{\sigma\sigma})
\ee
into the inner indices along a nontrivial loop. This way of creating $Z_2$ gauge twist lines is equivalent to using the string operators defined in Ref.\onlinecite{Levin2005}. Note that, similar to the SPT case discussed in the previous section, when inserting time reversal symmetry twist lines in the $x$ direction in the presence of a $Z_2$ gauge twist line in the $y$ direction, the inner symmetry operator at the crossing point needs to be changed to $M'_{\cT}$.

As $M_{\cT}$ and $M_{Z_2}$ are exactly the same as those used in the SPT case, the calculation of the composition between time reversal and $Z_2$ gauge twist lines seems to give exactly the same result as shown in section \ref{2DSPT_Z2Z2T}. One might want to conclude that there are correspondingly four SET phases. However, there is one major distinction between the SPT and the gauged version. In the gauged version, because the tensor remains invariant under the action of $M_{Z_2}$ on the inner indices, the inner symmetry operator induced by local time reversal symmetry action is not unique. In particular, we can redefine the inner symmetry operator for time reversal as
\be
\tilde{M}_{\cT} = M_{\cT}M_{Z_2}
\ee
When calculating the composition of twist lines, we need to do the calculation for both $M_{\cT}$ and $\tilde{M}_{\cT}$.

When $\alpha$ is trivial, i.e. we have the toric code topological order, $M_{\cT}$ and $\tilde{M}_{\cT}$ give the same result and we find two types of composition rules for the time reversal twist lines depending on the choice of $\eta$, similar to what we have seen in the SPT case. We interpret this difference in twist line composition as reflecting the $\cT^2=\pm 1$ transformation rule on the $Z_2$ gauge defects.

However, when $\alpha$ is nontrivial, i.e. we have the double semion topological order, $M_{\cT}$ and $\tilde{M}_{\cT}$ give different results. Consider the composition of two time reversal twist lines in the presence of an orthogonal $Z_2$ gauge twist line, similar to the configuration shown in Fig.\ref{fig:Z2Z2Tf_yxx}. When $\eta$ is trivial, calculation with $M_{\cT}$ yields a phase factor of $+1$ while that with $\tilde{M}_{\cT}$ gives $-1$. On the other hand, when $\eta$ is nontrivial, calculation with $M_{\cT}$ gives $-1$ while that with $\tilde{M}_{\cT}$ gives $+1$. Therefore, the distinction between the different choices of $\eta$ disappears. Indeed, there is only one SET phase with double semion type topological order and time reversal symmetry.\cite{Burnell2015}. The $\cT^2$ value is no longer well defined on the $Z_2$ gauge defects in the double semion state as they have semionic statistics and are not invariant under time reversal symmetry.

\section{Conclusion and open questions}
\label{sum}

In summary, we have proposed a way to `gauge' time reversal symmetry in the tensor network representation of many-body entangled quantum states. First we define a local action of time reversal symmetry in the tensor network states. The tensor network representation provides a natural way to divide the global wave function coefficient into local pieces. Based on this division, we can define the local action of complex conjugation, which is the key in defining local actions of anti-unitary symmetries. Then we can discuss how to introduce time reversal twists induced by time reversal fluxes through the nontrivial loops in the system. Moreover, by composing the time reversal twists, we can extract topological invariants of the phase from the projective composition rules. In particular, we demonstrate how this works for 1D time reversal SPT phases, which is a re-interpretation of the procedure used to determine the SPT order from the matrix product state representation of the state. Moreover, we studied a 2D time reversal symmetric state with trivial SPT order, 2D SPT states with $Z_2\times Z_2^T$ SPT order and 2D $Z_2$ gauge theories with time reversal symmetry. For the 2D time reversal symmetric state with trivial SPT order, we find that all the projective composition rules are trivial, as it should be for a trivial SPT state. For the 2D SPT states with $Z_2\times Z_2^T$ SPT order, the projective composition rules of the $Z_2$ and time reversal symmetry twist lines allow us to distinguish all four phases in the $\mathbb{Z}_2\times \mathbb{Z}_2$ classification. For the $Z_2$ gauge theories with time reversal symmetry, we find three different types of projective composition rules for the $Z_2$ gauge twist lines and the time reversal symmetry twist lines. This allows us to distinguish three SET phases with $Z_2$ gauge theory topological order and time reversal symmetry. Two of them have the toric code type topological order with the $Z_2$ gauge defect transforming as $\cT^2=1$ or $\cT^2=-1$ respectively. The third phase has double semion topological order and no further distinction can be made based on the time reversal transformation on the gauge defects\cite{Burnell2015}. 


This work just represents our first attempt at gauging time reversal symmetry and many questions remain. First of all, our discussion is totally based on quantum states. Is there a way to gauge time reversal symmetry on the Hamiltonian of the system? For unitary symmetries, we know how to do this. Of course, for time reversal symmetry, we can start from the gauged ground state and construct the corresponding gauged Hamiltonian. However, if we do not know the ground state, do we have a generic way to gauge the Hamiltonian? If we know how to do this, we can gauge time reversal symmetry not only in gapped systems, but in gapless systems as well.

Secondly, we ask if there is a \textit{dynamical} time reversal gauge theory. In this paper, we are only discussing non- dynamical configurations of time reversal twists. Can we promote it to a dynamical gauge theory by defining a time reversal gauge field and introducing quantum dynamics to it, as can be done for unitary symmetries like $Z_2$?

Moreover, we want to know what is the general procedure for determining time reversal protected and time reversal enriched topological orders using time reversal twists. In particular, we want to know how to identify time reversal related topological orders in 3D. In 3D, time reversal invariant SPT phases have a $\mathbb{Z}_2\times \mathbb{Z}_2$ classification. The group cohomology classification gives one $\mathbb{Z}_2$\cite{Chen2012a,Chen2013a} and the other $\mathbb{Z}_2$ is beyond the group cohomology classification\cite{Vishwanath2013}\cite{Kitaev_private}. One might expect, based our experience in 1D and 2D, that local time reversal operations in these 3D states leads to time reversal twist membranes, from which topological invariants of the SPT orders can be extracted. However, as we argue below, a time reversal twist membrane does not exist for the nontrivial phase in the beyond group cohomology $\mathbb{Z}_2$, at least not in the form we would expect. 


The nontrivial phase in the beyond group cohomology $\mathbb{Z}_2$ is special in how time reversal symmetry acts on the surface of the system. The system has half quantized surface thermal Hall effect.\cite{Vishwanath2013} That is, we can break time reversal symmetry in opposite ways on the left and right hand side of the surface and find a gapless edge state with chiral central charge $c_-=8$ between the gapped left and right half. Because time reversal symmetry on the surface maps the state on the left hand side to that on the right hand side, it maps between 2D states with different chirality. This can only be accomplished by a large quantum circuit with circuit depth scaling linearly with the size of the 2D surface. As we can see from all the other examples discussed in this paper, the inner symmetry operators induced by local symmetry actions on the tensors takes very similar form to the symmetry transformations on the edge / surface of the state. For all the other examples, the symmetry action on the edge / surface, hence the inner symmetry operators, are composed of a few layers of unitary operators. However, for the beyond group cohomology SPT state, the surface symmetry transformation cannot take this form. Therefore, we expect that the transformation induced by local symmetry action on the tensors cannot be described by simple inner symmetry operators on each inner index, hence complicating the discussion of time reversal twist membranes in the state.

For the within group cohomology $\mathbb{Z}_2$, a time reversal twist membrane can indeed be found. We leave it to future work to study how to extract topological invariants from it.

Finally, one may ask the question -- can gapped quantum ground states be generally represented as tensor network states? This is rigorously known to be true only in 1D \cite{Hastings2007}. The question of representing chiral 2D states with tensor networks was recently discussed\cite{Dubail2013, Wahl2013}, although the tensor networks have power law decaying correlation. Here we are interested in time reversal invariant, and hence non-chiral, phases. For example, can topological insulators and superconductors with time reversal symmetry be represented as short range correlated tensor network states? If so, will the gauged time reversal analysis discussed here give us a deeper understanding of these states?  We leave these interesting questions for future work.

After finishing this paper, we learned about the work of Kapustin\cite{Kapustin2014} where topological terms for time reversal gauge field is defined on non-orientable space time manifolds. 

\acknowledgments

XC wants to thank Xiao-Gang Wen, Alexei Kitaev, Lukasz Fidkowski, Max Metlitski and Mike Zaletel for very helpful discussions. XC is supported by the Miller Institute for Basic Research in Science at UC Berkeley, the Caltech Institute for Quantum Information and Matter and the Walter Burke Institute for Theoretical Physics. AV is supported by NSF DMR 0645691.


\end{document}